\documentclass[iop,apj]{emulateapj}
\usepackage{natbib}
\usepackage{graphicx}
\usepackage{epsfig}
\usepackage{subfigure}
\usepackage[twoside]{rotating}

\newcommand{\HII}{H~{\sc ii}}
\def\ca{$\mathcal{A}$}

\slugcomment{Accepted, June 30 2014}

\shorttitle{The MIR Structure of W49A}
\shortauthors{D. J. Stock et al.}

\begin{document}
\title{The mid-infrared appearance of the Galactic Mini-Starburst W49A.}

\author{D. J. Stock\altaffilmark{1}, E. Peeters\altaffilmark{1,2}, W. D.-Y. Choi\altaffilmark{1}, M. J. Shannon\altaffilmark{1}}
\altaffiltext{1}{Department of Physics and Astronomy, University of Western Ontario, London, ON, N6A 3K7, Canada}
\altaffiltext{2}{SETI Institute, 189 Bernardo Avenue, Suite 100, Mountain View, CA 94043, USA}

\begin{abstract}
The massive star forming region W49A represents one of the largest complexes of massive star formation present in the Milky Way and contains at least fifty young massive stars still enshrouded in their natal molecular cloud. We employ \textit{Spitzer}/IRS spectral mapping observations of the northern part of W49A to investigate the mid-infrared (MIR) spatial appearance of the polycyclic aromatic hydrocarbon (PAH) bands, PAH plateau features, atomic lines and continuum emission. We examine the spatial variations of the MIR emission components in slices through two of the ultra compact-\HII\ (UC-\HII) regions. We find that the PAH bands reproduce known trends, with the caveat that the 6.2 \micron\ PAH band seems to decouple from the other ionized PAH bands in some of the UC-\HII\ regions -- an effect previously observed only in one other object: the giant star forming region N66 in the LMC. Furthermore, we compare the nature of the emission surrounding W49A to that of `diffuse' sightlines. It is found that the surrounding emission can be explained by line of sight emission, and does not represent true `diffuse' material. Additionally, we examine the MIR appearance of star formation on various scales from UC-\HII\ regions to starburst galaxies, including a discussion of the fraction of PAH emission in the 8 \micron\ IRAC filter. We find that the MIR appearance of W49A is that of a starburst on large scales yet its individual components are consistent with other galactic \HII\ regions.
\end{abstract}

\keywords{HII regions; dust; extinction; ISM: individual objects W49A; infrared: ISM; galaxies: starburst}

\section{Introduction}
The mid-infrared (MIR) appearance of many astronomical objects is dominated by the unidentified infrared bands at 3.3, 6.2, 7.7, 8.6, 11.2, and 12.7 \micron\ (UIRs; c.f. \citealt{1973ApJ...183...87G}; \citealt{1989ApJ...341..278G}; \citealt{1989ApJ...341..246C}). The leading candidates as carriers of the bands are polycyclic aromatic hydrocarbons (PAHs), a family of molecules which consist of tiled aromatic carbon rings surrounded by hydrogen atoms on the periphery. PAHs are excited by UV radiation and then relax through  vibrational emission which make up the IR bands (\citealt{1989ARA&A..27..161P}; \citealt{1989ApJS...71..733A}). Variations in the relative strength of the individual PAH bands have long been known, both from source to source and within individual, spatially resolved sources (e.g. \citealt{1989ApJ...344..791B,1989ApJ...341..278G}). The advent of space based mid-IR observatories (e.g \textit{ISO}, \citealt{1996A&A...315L..27K}; and \textit{Spitzer}, \citealt{2004ApJS..154....1W}) has vastly increased the number of observations where variations are seen, spurring theoretical and experimental studies. It is clear now that there are many effects which can dramatically alter the band strengths of the MIR spectrum of an individual PAH, however the most important are ionization and molecular size \citep{1989ApJS...71..733A}. While these effects can also alter the precise shape of the bands (their profiles), such variations are generally confined to circumstellar material \citep{2002AA...390.1089P}.

Several studies revealed the existence of `families' of PAH bands whose relative strengths correlate well with each other, and poorly with other bands (e.g. \citealt{1996ApJ...460L.119J}; \citealt{2001A&A...370.1030H}; see \citealt{2008ARA&A..46..289T} for a complete list). The dominant family, comprising the 6.2, 7.7. and 8.6 \micron\ PAH bands is responsible for most of the MIR PAH emission. It was then shown that this family is generated by ionized PAH molecules (e.g. \citealt{1985ApJ...290L..25A, 1993ApJ...408..530D, 1999ApJ...511L.115A} and references therein). The other dominant bands, at 3.3 and 11.2 \micron\ were attributed to neutral PAH emission.

The `requirement' that a UV photon be absorbed to emit the MIR PAH bands leads to the majority of galactic PAH emission tracing the sites of massive star formation. However, the PAH emission does not emanate from the ionized \HII\ regions which are most intimately associated with the ionizing stars as the environment is too hostile. The PAH molecules exist in the photodissociation regions (PDRs) which form in the surroundings of the ionized regions where the UV field has been depleted of the highest energy photons responsible for PAH destruction. This PAH emission is routinely observed in both Galactic (e.g. \citealt{2000A&A...357.1013V,  2001A&A...370.1030H, 2001A&A...372..981V, 2007ApJ...660..346P}) and extra-galactic contexts (e.g. \citealt{2002A&A...382.1042V, 2006ApJ...653.1129B, 2007ApJ...656..770S, 2008ApJ...676..304B, 2008ApJ...682..336G}). There is a fundamental difference between the two groups of studies though, in that the spatial scales probed are wildly different. An understanding of the link between the large unresolved star forming regions which dominate in external galaxies, especially starburst galaxies, and the less massive star formation complexes present in the Galaxy is required to calibrate the properties of external galaxies. Efforts to understand the connection between these regimes in terms of the PAH emission have been limited so far, with \citet{2004ApJ...613..986P} being the first to consider both mass and star formation activity regimes concurrently. 

To better characterize this connection, we analyze the MIR appearance of the galactic star forming region W49A\footnote{W49A is now seen to consist of three distinct regions: W49A north (or sometimes northwest); W49A south; and W49A southwest (e.g. \citealt{1977ApJ...211..786H}). \citet{1993ApJ...413..571S} note that W49A has come to be synonymous with W49A-N; we will follow this usage and refer to W49A-N as W49A.}, focusing on the properties of the PAH emission. W49A is one of the largest and most active sites of massive star formation in the galaxy (e.g. \citealt{2003Msngr.114...35A}; \citealt{2010A&A...520A..84P}; \citealt{2012A&A...542A...6N}). W49A consists of up to a hundred massive O type stars, many still encased in their natal ultra~compact~\HII\ regions (UC~\HII; \citealt{2002ARA&A..40...27C}) surrounded by copious molecular cloud material (\citealt{2010A&A...520A..84P}). The entire complex lies at a distance of 11.4 kpc \citep{1992ApJ...393..149G}.

In order to investigate this relationship within a single object, archival \textit{Spitzer}/IRS spectral maps of W49A and its associated UC~\HII\ regions are investigated in order to probe the PAH band emission properties in addition to the overall MIR appearance. This paper is organized as follows: Section~\ref{sec:dra} presents descriptions of the observations along with the data reduction, extinction and the MIR decomposition. In Section~\ref{sec:hii}, the spectra of the various UC~\HII\ regions are presented, along with discussion of their individual characteristics. Subsequently, maps of the PAH emission are presented and discussed in Section~\ref{sec:maps} along with traditional PAH correlation studies in Section~\ref{sec:corrs}. Finally, discussion of our results in terms of the structure of UC~\HII\ regions, diffuse emission and extragalactic star formation, including the IRAC 8 \micron\ SFR indicator, is given in Section~\ref{sec:discuss}, followed by a summary of our conclusions in Section~\ref{sec:conc}.

\section{Observations, Data Reduction and Analysis}\label{sec:dra}

\begin{figure*}
	\begin{center}
	\includegraphics[width=16cm]{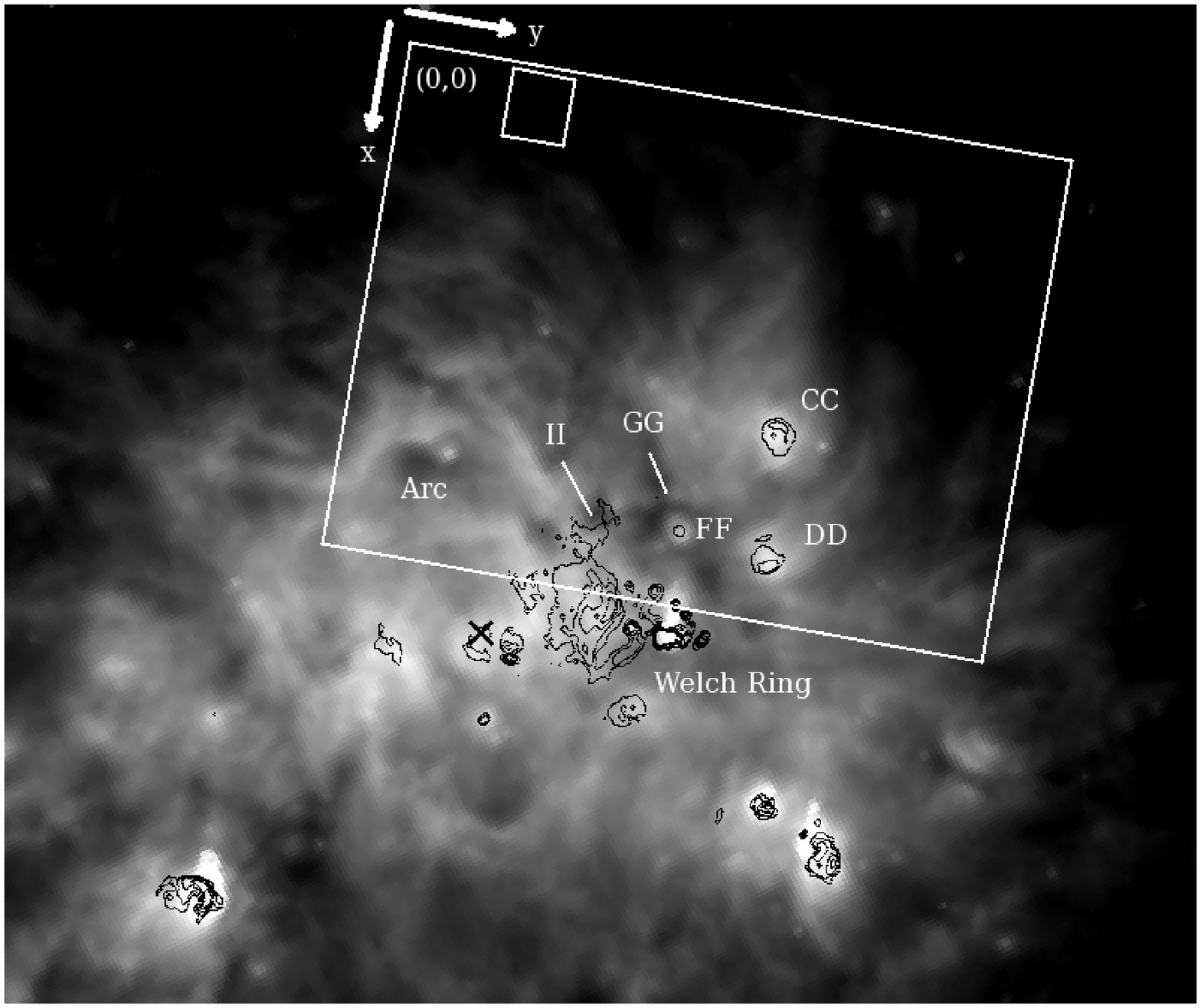}
	\end{center}
	\caption{IRAC [8.6] image of W49A using GLIMPSE data (\citealt{2003PASP..115..953B,2009PASP..121..213C}); with \textit{Spitzer}/IRS field of view indicated (large white square, origin and axes indicated) and 3.6 cm radio emission (black contours; \citealt{1997ApJ...482..307D}). The main UC~\HII\ regions CC, DD, EE, GG, II are indicated (following the naming convention of \citealt{1984ApJ...283..632D} and \citealt{1997ApJ...482..307D}). The central cluster is marked with a black cross and the position of the Welch ring is indicated. The small white box represents the extraction region for the diffuse emission. North is up and east is to the left. The coordinates of the (0, 0) pixel of the IRS cube are: 19h 10m 18s +9$^\circ$ 8" 48' with the $x$ axis of the cube having a position angle of $\sim$ 80$^\circ$ E of N. }
	\label{fig1}
\end{figure*}

\subsection{Observations} W49A was observed with the Short-Low (SL) module of the Infrared Spectrometer (IRS; \citealt{2004ApJS..154...18H}) on-board the \textit{Spitzer} Space Telescope (\citealt{2004ApJS..154....1W}; PID: 63, AOR: 16206848). The first (SL1, 7.4 - 14.5 \micron, 1.8\arcsec/pixel) and the second (SL2, 5.2 - 7.7 \micron, 1.8\arcsec/pixel) order spectra of the IRS-SL module were used to create a spectral map, with spectral resolutions $^\lambda/_{\delta\lambda}$ between 64 and 128. These observations cover an area of roughly 3\arcmin\ $\times$ 2\arcmin\ (Figure~\ref{fig1}). The point spread function (PSF) is about 2 pixels, roughly 3.6\arcsec.

The alignment of the map relative to W49A (shown in Figure~\ref{fig1}) is centered upon the northern section of W49A and includes at least five of the UC~\HII\ sources identified by \citet{1997ApJ...482..307D} along with a section of the expanding shells, labeled as the `arc' region, and the northern extremities of the Welch ring.

\subsection{Data Reduction} The spectral mapping data were processed through the pipeline reduction software at the \textit{Spitzer} Science Center (Version 18.18). The BCD images were cleaned using CUBISM \citep{2007PASP..119.1133S}, in which any rogue or otherwise `bad' pixels were masked, firstly using the built in `AutoGen Global Bad Pixels' with settings `Sigma-Trim' = 7, `MinBad-Frac' = 0.5 and `AutoGen Record Bad Pixels' with settings `Sigma-Trim' = 7 and `MinBad-Frac' = 0.75 and subsequently by manual inspection of each cube. Bogus data at the edges of the slit were eliminated by applying a wavesamp of 0.06 -- 0.94. 

Finally, the SL2 spectra were scaled to match the SL1 spectra in order to correct for the mismatches between the two order segments. The bonus order SL3 was used to determine the scaling factors, since it overlaps slightly with both SL1 and SL2. These scaling factors typically of the order of 10\%, in agreement with the typical values of 10\% obtained by \citet{2007ApJ...656..770S}. In order to correct for the mis-matched pixel (1.8\arcsec\ $\times$ 1.8\arcsec) and beam (3.6\arcsec\ diameter) sizes, the final spectral map was created by averaging the spectra in a 2$\times$2 aperture for each point in the map and then regridding the map such that the pixel centers correspond to the centers of the 2$\times$2 apertures. This results in a map with overlapping extraction apertures such that only every second pixel in each direction is an independent measurement. Throughout the rest of the work only independent pixels were used in the analysis. At the distance of W49A (11.4 kpc), each independent 2$\times$2 pixel area in the cube represents a physical area of 0.2 pc $\times$ 0.2 pc.

\begin{figure*}
	\begin{center}
	\includegraphics[width=14cm]{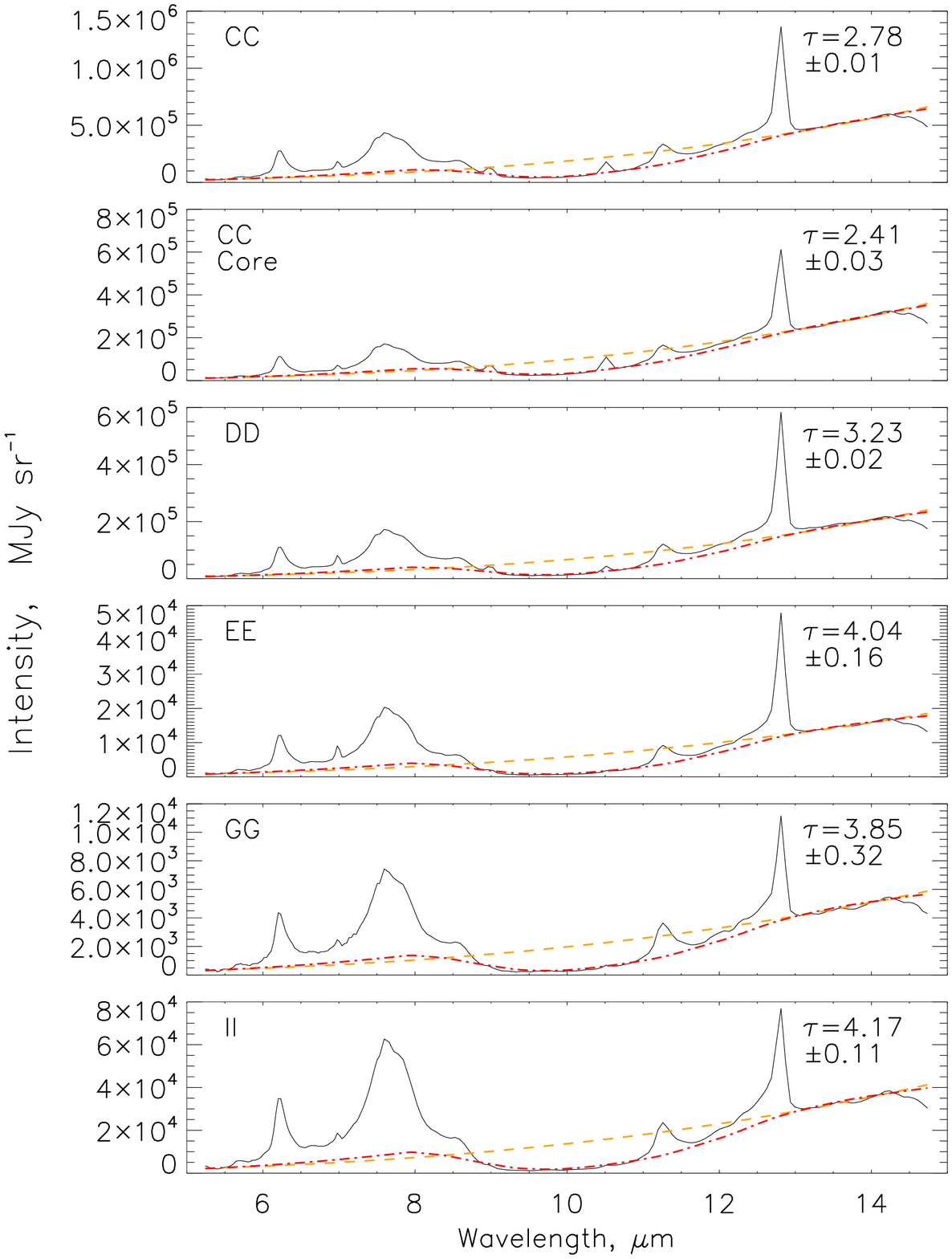}
	\end{center}
	\caption{Integrated spectra for different regions of W49A. Silicate optical depths are indicated for each spectrum. The orange dashed line represents the Spoon method continuum (see \citet{2007ApJ...654L..49S}). The red dot-dashed line represents the final continuum given by the iterative Spoon method (as described by \citealt{2013ApJ...771...72S}) for each case, after the application of the \citet{2006ApJ...637..774C} extinction law.}
	\label{fig:spec}
\end{figure*}

\subsection{Extinction}\label{sec:ext}
\begin{figure}
	\begin{center}
 	\includegraphics[width=7.5cm]{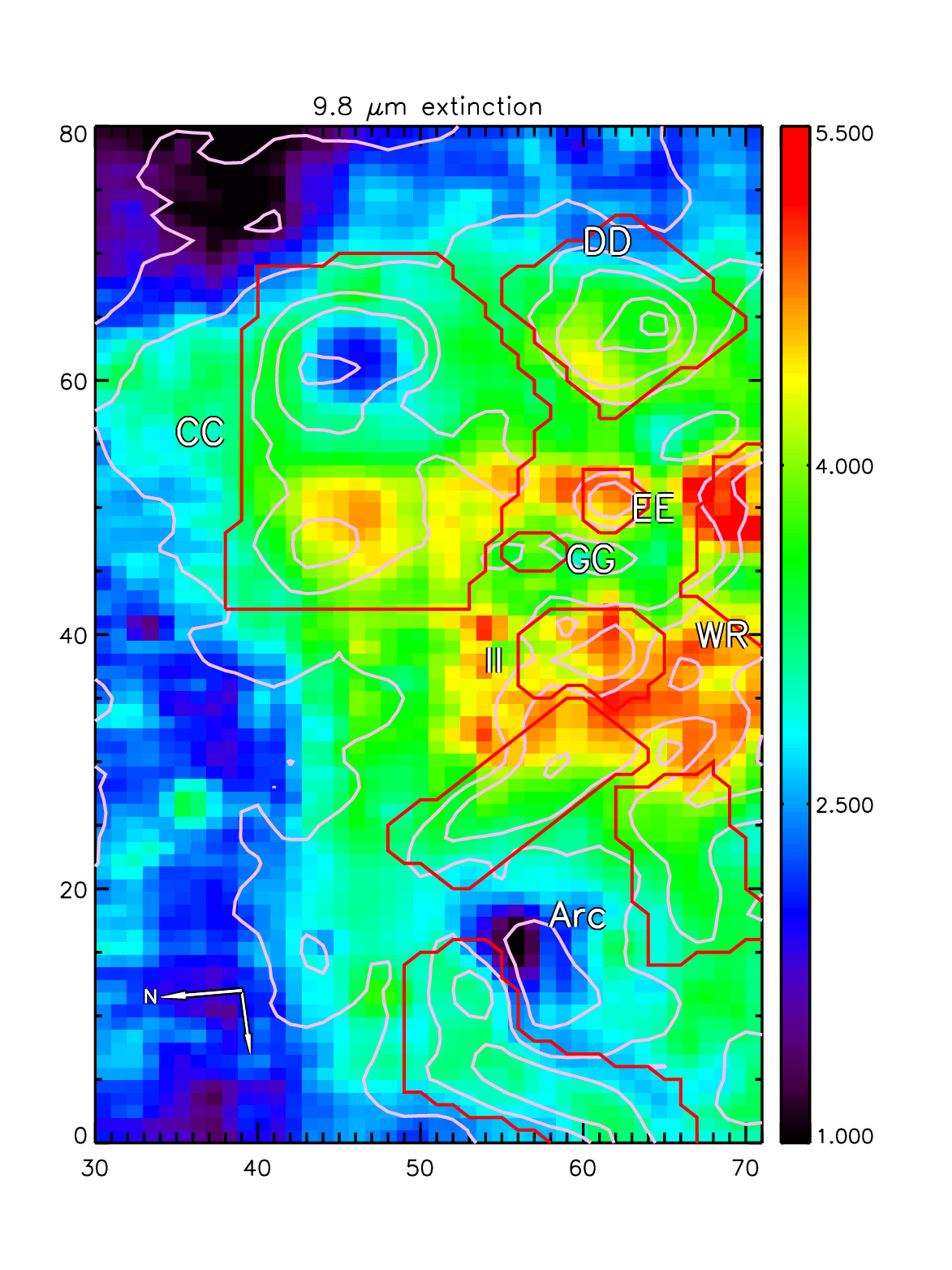}
	\end{center}
	\caption{Map of extinction caused by 9.8 \micron\ silicate absorption expressed as the K band extinction in magnitudes (A$_K$) across W49A derived using the iterative Spoon method \citep{2013ApJ...771...72S}. Pink contours represent 7.7 \micron\ PAH fluxes (see Section~\ref{sec:maps}) red boxes indicate the extraction regions used for the different UC~\HII\ regions and the arc segments.}
	\label{fig:ext}
\end{figure}

The integrated MIR spectra for each of the various UC-\HII\ regions in the field of view is shown in Figure~\ref{fig:spec}. Each of the spectra clearly show the influence of silicate absorption to various degrees. To investigate the effects of extinction on the entire IRS field of view, the iterative Spoon method as described by \citet{2013ApJ...771...72S} -- based upon the method of \citet{2007ApJ...654L..49S} -- was used to measure the optical depth of the silicate feature for each pixel. The optical depths were then transformed into K band extinctions using the \citet{2006ApJ...637..774C} extinction law. The resulting map of silicate absorption is shown in terms of the K band extinction, A$_K$, in Figure~\ref{fig:ext}.

The method described by \citet{2013ApJ...771...72S} extends the Spoon method to compensate for extinction in the 14-16 \micron\ region by introducing an iterative reapplication of the Spoon method to the residual Silicate absorption in a partially dereddened spectrum until there is no residual. The final continuum is then compared to the observed flux at 9.8 \micron\ to calculate the total optical depth at 9.8 \micron\ ($\tau_{9.8}$). This figure can then be converted into A$_K$ using a small correction factor (A$_K$ = 1.079 $\times$ $\tau_{9.8}$; \citealt{2013ApJ...771...72S}). The numerical uncertainties associated with this technique are calculated using standard propagation of uncertainty and are quoted where appropriate (e.g. Figure~\ref{fig:spec}). As one might expect, the numerical uncertainty is highest for the sources with the lowest flux, as the uncertainty of the flux of the 9.8 \micron\ region dominates the overall uncertainty. In terms of the uncertainties of the map presented in Figure~\ref{fig:ext} we find that the numerical uncertainty is around 0.5 magnitudes for the highest extinction regions and 0.1 -- 0.2 magnitudes for the lowest extinction regions. However there are also clear systematic uncertainties associated with the iterative method. The reliance upon the \citet{2006ApJ...637..774C} extinction law to deredden the spectra assumes that the \citet{2006ApJ...637..774C} extinction law is universally applicable with no variations across the map. To test this hypothesis, in Figure~\ref{fig:spec} we have included both the initial Spoon method continuum (orange dashed) and the final iterative Spoon method continuum extinguished using the \citet{2006ApJ...637..774C} extinction law and the final optical depth (red dot-dash). The second quantity can be thought of as the appearance of the emitted power law continuum after the effects of extinction are applied. In each case, the shape of the power law continuum after the application of the \citet{2006ApJ...637..774C} extinction law matches the observed shape of the continuum, implying that the \citet{2006ApJ...637..774C} extinction law is close to that seen in W49A.

As silicate absorption can have a large effect on the fluxes of the PAH features and atomic lines, each spectrum in the map was dereddened using the \citet{2006ApJ...637..774C} extinction law, creating an extinction corrected cube. If one takes the opposing stategy, and merely corrects the fluxes in the observed cube using the average extinction for each band the differences are around 10\%, while for the coadded spectra given in Table~\ref{table:PAHints} this can rise to as much as 20-30\% depending on how much variability there is in the extinction of the coadded pixels. This difference arises from the sometimes steep gradients in the extinction curve with respect to wavelength in the 5--15 \micron\ region. It should be noted that the [S~{\sc iv}] 10.5 \micron\ line is heavily affected by the presence of the silicate absorption feature so our later choices of ionization proxy will be sensitive to the details of the extinction law and dereddening methods applied.

\subsection{Spectral Decomposition}\label{sec:decomp}

\begin{figure}
	\begin{center}
	\includegraphics[width=7.5cm]{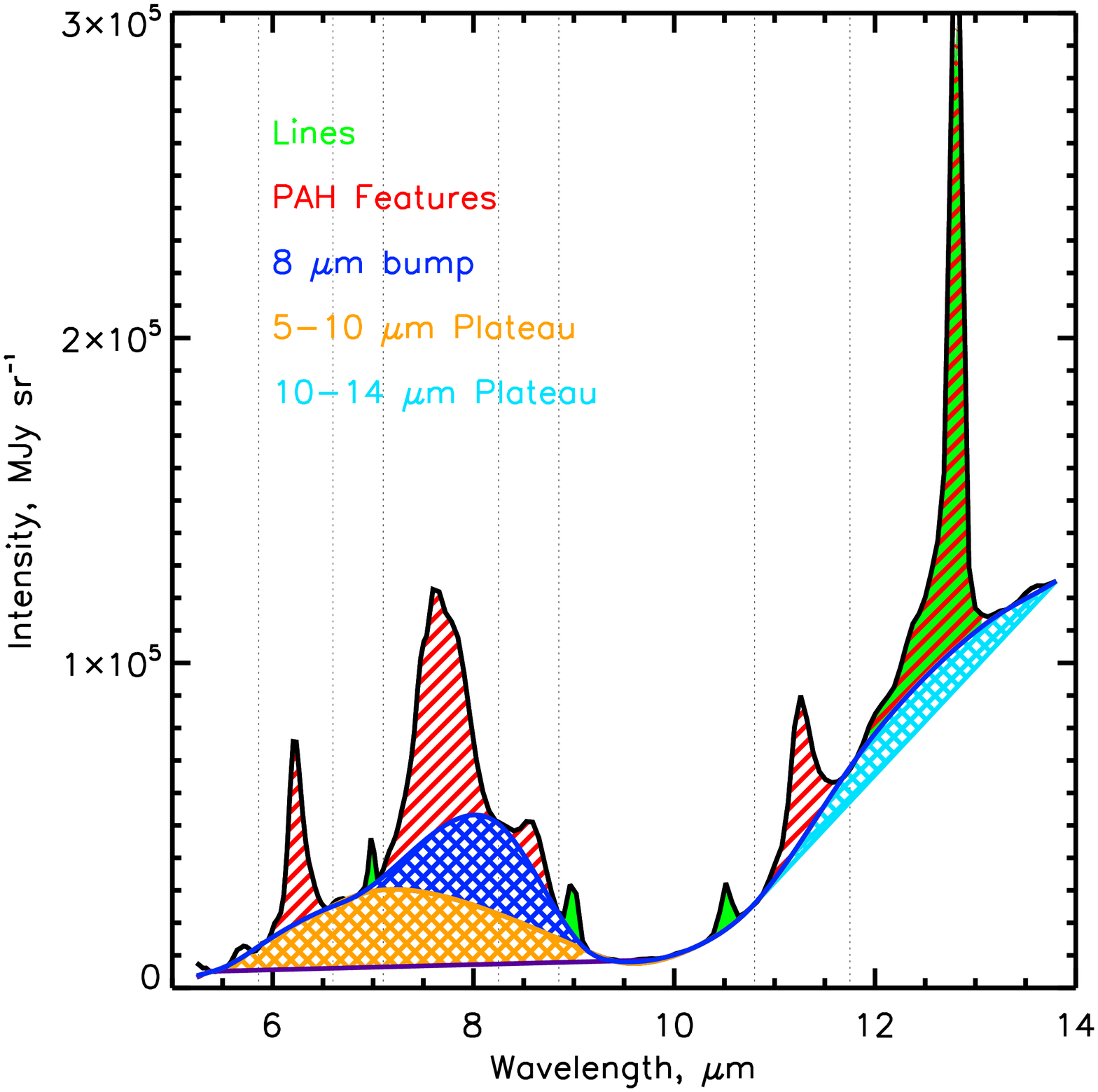}
	\end{center}
	\caption{Example of the spectral decomposition using the spline method for the integrated spectrum of the arc region. The 5--10 \micron\ plateau, 10--14 \micron\ plateau and 8 \micron\ bump are cross hatched orange, turquoise, and blue respectively, the PAH features are hatched red and the lines are colored green. The 12.7 \micron\ region blend is colored red and green as it is a mixture of line and PAH emission. The primary continuum used throughout this work is represented by the blue line and the secondary continuum used to define the 5--10 \micron\ plateau is represented by the orange line. Dotted lines show the integration ranges for the main PAH features.}
	\label{fig:decomp}
\end{figure}

Example spectra from the regions of W49A which are discussed in this paper are shown in Figure~\ref{fig:spec}. The decomposition of each MIR spectrum into distinct features and their subsequent measurement was performed using the so called `spline method' (e.g. \citealt{2000A&A...357.1013V, 2001A&A...370.1030H,2002AA...390.1089P, 2004ApJ...611..928V, 2010A&A...511A..32B}, Peeters et al. 2014 (in prep)), which we summarize here. An example of the spline method as applied to the integrated spectrum of the arc region is shown in Figure~\ref{fig:decomp}. The continuum was removed from each spectrum by computing a local spline polynomial fit to the continuum (including a point at 8.2 \micron) and then subtracting the spline. 

The area under the well known PAH bands in the 5--10 \micron\ regions was split into two distinct regions (following, for example, \citealt{2010A&A...511A..32B}). Initially, the `5--10 \micron\ plateau' was defined by computing the area between a secondary continuum spline (in which the 8.2 \micron\ point was not included) and the line connecting the observed fluxes at 5.5 and 10 \micron\ respectively. The second component, which we name the `8 \micron\ bump', was computed by integrating the difference between the primary spline and the secondary spline - tracing the area directly under the 7.7 and 8.6 \micron\ PAH bands. The longer wavelength plateau was measured by integrating the result of subtracting a spline continuum fitted to points at around 10 \micron\ and 14 \micron\ from the main continuum spectrum over the range from 10-14 \micron.

Measurement of the PAH features were performed in two distinct ways. For the strong features, at 6.2, 7.7, 8.6 and 11.2 \micron, all of the flux above the continuum associated with the feature was summed. For the weaker bands, at 6.0, 11.0, 12.0, 12.7 \micron\ etc,  various fitting routines have been developed to reliably measure their flux. Where the weaker features overlapped with stronger features (e.g. 6.0/6.2, 11.0/11.2) gaussian fits to the weaker feature were obtained and subtracted from the total integrated intensity to correct the flux of the stronger feature (see \citealt{2010ApJ...718..558A}, Peeters et al., 2013).

The most prominent blend, in the 12--13 \micron\ region, involves the 12.0 \micron\ PAH band, 12.3 \micron\ H$_2$ line, 12.7 \micron\ PAH band and 12.8 \micron\ [Ne~{\sc II}] line. The profile of the 12.7 \micron\ PAH feature is known to be strongly asymmetric \citep{2001A&A...370.1030H}. While the 12.7 \micron\ feature is known to vary in profile from source to source, it has been found to be remarkably consistent in profile for \HII\ regions and ISM material \citep{2001A&A...370.1030H}. In order to remove the 12.7 \micron\ component of the blend, the average ISM 12.7 \micron\ band shape (given by \citet{2001A&A...370.1030H}, measured using ISO-SWS) was downsampled to the resolution of the IRS-SL spectra and subtracted from the blends after being scaled to match the intensity of the emission in the 12.4 to 12.6 \micron\ region. The flux in the 12.7 \micron\ band was found by integrating all of the flux in the scaled 12.7 profile. Subtracting the 12.7 \micron\ band from the observed spectrum left the 12.3 \micron\ H$_2$ line and the 12.8 \micron\ [Ne~{\sc II}] line on a flat continuum. Each line was fitted with a gaussian of FWHM set by the instrumental resolution. 

The systematic uncertainty associated with our decomposition of the 12.7/12.8 blend was estimated by applying the method to artificial spectra. The mean ISM spectrum measured by \citet{2001A&A...370.1030H} was used as a basis for the 12.7 PAH band, however it was again downsampled from the resolution of ISO-SWS to that of IRS-SL. The [Ne~{\sc ii}] 12.8 \micron\ line is constructed with a gaussian of varying peak amplitudes and width set by the IRS/SL spectral resolution. This artificial blend is then downsampled to the instrumental resolution of \textit{Spitzer}/IRS SL. We apply our decomposition method, which returns measurements of the integrated fluxes of the 12.7 \micron\ PAH band and the 12.8 \micron\ [Ne~{\sc ii}] line. We can then compare our measured PAH flux to the input PAH flux as a function of varying [Ne~{\sc ii}] line strengths. It was found that the choice of the 12.4 to 12.6 \micron\ window eliminated potential contamination from the 12.8 \micron\ [Ne~{\sc ii}] line, while if the upper limit of the window was increased to 12.7 \micron, a substantial contamination occurred of around 40\% at the maximum [Ne~{\sc ii}] line strength for these observations.

The other atomic lines present in the 5-14 \micron\ range were each fitted with a gaussian with FWHM set by the instrumental resolution. Aside from the 12.8 \micron\ [Ne~{\sc ii}] line and the 12.3 \micron\ H$_2$ line, the other lines ([Ar~{\sc ii}] 6.98 \micron, [Ar~{\sc iii}] 8.99 \micron, H$_2$ 9.6 \micron\ and [S~{\sc iv}] 10.5 \micron) all appear in relative isolation or on top of an even broader feature that can be closely approximated by using a polynomial fit to the local continuum. 

For the gaussian fits, uncertainties are generated in the fitting process, while for the features for which the integration was used, uncertainties were calculated by measuring the RMS noise in various wavelength windows and then combining this with each flux measurement to find the signal to noise ratio. The technique of comparing the flux to the measured RMS noise was also used for the 12.7 \micron\ feature uncertainty. 

Several other more complex methodologies have been proposed and implemented for the measurement of the PAH bands, each with its strengths, weaknesses and assumptions (PAHFIT; \citealt{2007ApJ...656..770S}, PAHTAT; \citealt{2012A&A...542A..69P} and AMES PAH database; \citealt{2010ApJS..189..341B}). \citet{2007ApJ...656..770S} and \citet{2008ApJ...679..310G} showed that the behavior of the PAH bands was independent of the method of measurement (while the obtained fluxes do depend on the method). Hence, we are confident that the decomposition described here will not bias the results significantly.

\section{Results}

\subsection{The UC~\HII\ Regions and their MIR IRS spectra}\label{sec:hii}
High angular resolution radio observations of W49A have revealed a wealth of embedded structure, with 45 continuum and water maser sources reported by \citet{1997ApJ...482..307D, 2000ApJ...540..308D}. Of these, several overlap with the field of view of the \textit{Spitzer}/IRS cube. The UC~\HII\ regions (radio continuum sources) CC, DD, EE, GG and II lie completely within the field of view. These UC~\HII\ regions lie just to the north of the `Welch Ring' (\citealt{1987Sci...238.1550W}, labeled in Figure~\ref{fig1} and later in Section~\ref{sec:maps} as  `WR') and northwest of what is thought to be the first stars to form in W49A (the central cluster; \citealt{2003ApJ...589L..45A, 2005A&A...430..481H}; also labeled in Figure~\ref{fig1}). 

The UC~\HII\ regions F, I, J and M lie close to the southern edge of the IRS cube. These UC~\HII\ regions are either part of, or immediately adjacent to, the Welch ring. They are much smaller and closer together than those which dominate the field of view and represent only a handful of pixels on the \textit{Spitzer}/IRS maps, so they shall not be considered in detail.

The UC~\HII\ regions in W49A comprise a variety of the morphological classes defined by \citet{1989ApJS...69..831W}. The UC~\HII\ regions CC and II are listed as `Shell' type, while EE and GG are unresolved.

In Figure~\ref{fig:spec}, the integrated spectra of the UC~\HII\ regions are shown. Each displays a typical \HII\ region spectrum, with strong [Ne~{\sc ii}] 12.8 \micron\ emission, silicate absorption and a class \ca\ PAH emission spectrum \citep{2002AA...390.1089P}. Weaker ionized line emission from [Ar~{\sc ii}] 6.99, [Ar~{\sc iii}] 8.99 and [S~{\sc iv}] 10.5 \micron\ is present in each of the spectra to varying degrees (strongest in CC/DD, weakest in GG). The integrated spectra also contain a selection of weaker PAH features at 5.7, 6.0, 11.0, 12.0, 13.5 and 14.2 \micron\ and weak H$_{2}$ emission at 9.6 and 12.3 \micron. 

In Table~\ref{table:PAHints}, the measured fluxes for the integrated spectra are shown, both in their observed form (i.e. measurements of the spectra in the Figure~\ref{fig:spec}) and after the dereddening process which will be discussed in Section~\ref{sec:ext}.

\begin{sidewaystable*}
\vspace{5cm}
\begin{center}
\caption{Fluxes and Characteristics of Features Detected in Integrated Spectra}\label{table:PAHints}

\medskip

\begin{tabular}{l r r r r r r r r }
Feature & CC & CC East & CC Core & DD & EE & GG & II & arc \\
\hline
\multicolumn{9}{c}{\textit{General Characteristics}}\\
                      $\tau_{9.8}$  &      2.78 &      3.35 &      2.41 &      3.23 &      4.04 &      3.85 &      4.17 &      2.92 \\
                          Area$^a$  &   1577.88 &    913.68 &    599.40 &    505.44 &    226.80 &     51.84 &     55.08 &   2831.76 \\
\hline
\multicolumn{9}{c}{\textit{Observed Fluxes$^b$}}\\
           I(6.0)  &      3.40 $\pm$      2.64 &      1.12 $\pm$      0.94 &      0.50 $\pm$      0.36 &      0.51 $\pm$      3.64 &      0.02 $\pm$      0.01 &      0.00 $\pm$      0.00 &      0.02 $\pm$      0.01 &      7.18 $\pm$      3.64 \\
           I(6.2)  &   1112.09 $\pm$      4.56 &    364.00 $\pm$      1.63 &    164.13 $\pm$      0.62 &    144.12 $\pm$      6.29 &      7.13 $\pm$      0.02 &      0.59 $\pm$      0.00 &      5.03 $\pm$      0.02 &   2228.65 $\pm$      6.29 \\
           I(7.7)  &   2699.28 $\pm$     10.25 &    899.99 $\pm$      3.66 &    388.39 $\pm$      1.39 &    347.31 $\pm$     14.13 &     19.73 $\pm$      0.04 &      1.72 $\pm$      0.01 &     15.62 $\pm$      0.04 &   5519.63 $\pm$     14.13 \\
           I(8.6)  &    240.30 $\pm$      6.49 &     68.51 $\pm$      2.31 &     41.77 $\pm$      0.88 &     28.96 $\pm$      8.94 &      1.08 $\pm$      0.03 &      0.09 $\pm$      0.00 &      0.60 $\pm$      0.02 &    468.41 $\pm$      8.94 \\
          I(11.0)  &      2.68 $\pm$      0.76 &      0.65 $\pm$      0.27 &      0.55 $\pm$      0.10 &      0.24 $\pm$      1.05 &      0.00 $\pm$      0.00 &      0.00 $\pm$      0.00 &      0.01 $\pm$      0.00 &      3.48 $\pm$      1.05 \\
          I(11.2)  &    368.56 $\pm$      7.20 &    113.71 $\pm$      2.57 &     58.30 $\pm$      0.98 &     43.31 $\pm$      9.95 &      1.63 $\pm$      0.03 &      0.18 $\pm$      0.00 &      1.16 $\pm$      0.03 &    752.19 $\pm$      9.95 \\
          I(12.0)  &     12.18 $\pm$      5.92 &      3.65 $\pm$      2.00 &      1.96 $\pm$      0.80 &      1.53 $\pm$      8.16 &      0.06 $\pm$      0.02 &      0.01 $\pm$      0.00 &      0.04 $\pm$      0.02 &     26.84 $\pm$      8.16 \\
          I(12.7)  &    469.84 $\pm$      7.25 &    144.22 $\pm$      2.59 &     74.14 $\pm$      0.99 &     60.48 $\pm$      9.99 &      2.73 $\pm$      0.03 &      0.21 $\pm$      0.00 &      1.85 $\pm$      0.03 &   1017.73 $\pm$      9.99 \\
 I([Ar~{\sc ii}])  &      6.06 $\pm$      2.09 &      2.15 $\pm$      0.75 &      0.77 $\pm$      0.28 &      1.04 $\pm$      2.88 &      0.05 $\pm$      0.01 &      0.00 $\pm$      0.00 &      0.01 $\pm$      0.01 &      9.22 $\pm$      2.88 \\
I([Ar~{\sc iii}])  &     12.78 $\pm$      3.74 &      2.75 $\pm$      1.34 &      2.74 $\pm$      0.51 &      1.13 $\pm$      5.16 &      0.01 $\pm$      0.02 &      0.00 $\pm$      0.00 &      0.00 $\pm$      0.01 &     21.07 $\pm$      5.16 \\
  I([S~{\sc iv}])  &     16.83 $\pm$      4.19 &      3.08 $\pm$      1.49 &      3.93 $\pm$      0.57 &      0.93 $\pm$      5.77 &      0.01 $\pm$      0.01 &      0.00 $\pm$      0.00 &      0.00 $\pm$      0.01 &     15.94 $\pm$      5.77 \\
 I([Ne~{\sc ii}])  &     91.95 $\pm$      4.19 &     27.92 $\pm$      1.49 &     14.69 $\pm$      0.57 &     13.12 $\pm$      5.77 &      0.47 $\pm$      0.02 &      0.02 $\pm$      0.00 &      0.13 $\pm$      0.01 &    172.07 $\pm$      5.77 \\
                  Cont. 14 \micron  &   2941.92 &    674.46 &    604.67 &    345.83 &     11.95 &      0.86 &      6.45 &   4924.13 \\
\hline
\multicolumn{9}{c}{\textit{Extinction Corrected Fluxes$^b$}}\\
           I(6.0)  &     12.47 $\pm$     19.69 &      4.56 $\pm$      7.88 &      1.56 $\pm$      2.08 &      1.78 $\pm$     32.90 &      0.10 $\pm$      0.16 &      0.01 $\pm$      0.03 &      0.10 $\pm$      0.13 &     26.36 $\pm$     32.90 \\
           I(6.2)  &   3878.71 $\pm$     34.00 &   1440.96 $\pm$     13.60 &    470.48 $\pm$      3.60 &    511.32 $\pm$     56.79 &     35.29 $\pm$      0.28 &      2.27 $\pm$      0.05 &     24.43 $\pm$      0.23 &   7808.72 $\pm$     56.79 \\
           I(7.7)  &   8892.69 $\pm$     76.41 &   3275.88 $\pm$     30.58 &   1094.26 $\pm$      8.09 &   1176.15 $\pm$    127.66 &     88.20 $\pm$      0.63 &      6.06 $\pm$      0.11 &     66.93 $\pm$      0.52 &  18480.55 $\pm$    127.66 \\
           I(8.6)  &   1376.92 $\pm$     48.33 &    479.16 $\pm$     19.34 &    186.98 $\pm$      5.11 &    178.52 $\pm$     80.74 &     10.65 $\pm$      0.40 &      0.57 $\pm$      0.07 &      6.27 $\pm$      0.33 &   2776.16 $\pm$     80.74 \\
          I(11.0)  &     50.11 $\pm$      5.69 &     17.50 $\pm$      2.28 &      6.84 $\pm$      0.60 &      5.80 $\pm$      9.51 &      0.29 $\pm$      0.05 &      0.02 $\pm$      0.01 &      0.25 $\pm$      0.04 &     81.24 $\pm$      9.51 \\
          I(11.2)  &   4293.62 $\pm$     53.40 &   1662.88 $\pm$     21.39 &    478.98 $\pm$      5.64 &    524.62 $\pm$     89.41 &     36.01 $\pm$      0.44 &      2.42 $\pm$      0.07 &     24.88 $\pm$      0.37 &   8517.22 $\pm$     89.41 \\
          I(12.0)  &     69.46 $\pm$     41.86 &     25.89 $\pm$     16.75 &      8.25 $\pm$      4.43 &      9.42 $\pm$     69.93 &      0.61 $\pm$      0.35 &      0.05 $\pm$      0.06 &      0.46 $\pm$      0.29 &    157.33 $\pm$     69.93 \\
          I(12.7)  &   2369.41 $\pm$     54.04 &    885.60 $\pm$     21.62 &    280.27 $\pm$      5.72 &    320.67 $\pm$     90.28 &     23.21 $\pm$      0.45 &      1.29 $\pm$      0.07 &     16.15 $\pm$      0.37 &   5258.31 $\pm$     90.28 \\
 I([Ar~{\sc ii}])  &     20.38 $\pm$     15.57 &      8.27 $\pm$      6.23 &      2.00 $\pm$      1.65 &      3.68 $\pm$     26.01 &      0.25 $\pm$      0.13 &      0.01 $\pm$      0.02 &      0.05 $\pm$      0.11 &     29.90 $\pm$     26.01 \\
I([Ar~{\sc iii}])  &     94.28 $\pm$     27.90 &     29.92 $\pm$     11.16 &     14.25 $\pm$      2.95 &     11.68 $\pm$     46.61 &      0.18 $\pm$      0.23 &      0.00 $\pm$      0.04 &      0.04 $\pm$      0.19 &    170.90 $\pm$     46.61 \\
  I([S~{\sc iv}])  &    180.56 $\pm$     31.19 &     52.16 $\pm$     12.48 &     30.24 $\pm$      3.30 &     16.45 $\pm$     52.11 &      0.15 $\pm$      0.20 &      0.01 $\pm$      0.03 &      0.09 $\pm$      0.19 &    201.94 $\pm$     52.11 \\
 I([Ne~{\sc ii}])  &    484.57 $\pm$     31.20 &    182.66 $\pm$     12.48 &     56.26 $\pm$      3.30 &     78.23 $\pm$     52.12 &      4.21 $\pm$      0.26 &      0.11 $\pm$      0.04 &      1.15 $\pm$      0.21 &    900.77 $\pm$     52.12 \\
                  Cont. 14 \micron  &  14363.16 &   4321.49 &   2324.73 &   2123.51 &    108.70 &      5.67 &     59.91 &  27192.44 \\

\end{tabular}

\bigskip
$^a$: in $\Box$\arcsec. $^b$: in 10$^{-13}$ W m$^{-2}$ (\micron$^{-1}$ for continua) \\
\end{center}

\end{sidewaystable*}

\subsubsection{Comments on Individual Regions}

\paragraph{CC} \citet{1997ApJ...482..307D} discussed the status of the UC~\HII\ regions CC, DD and EE based on their radio observations. They noted that CC possessed the strangest morphology (a broken shell, R = 0.34 pc) and discussed the idea that the extended emission to the east of CC is actually an outflow from the ionized region. This idea found support in the NIR observations of \citet{2004ASPC..322..365H}, who found that the NIR morphology matched that seen in the radio - the extended emission to the east and a corresponding gap in the shell around source CC.  \citet{2004ASPC..322..365H} also suggest that region CC is formed by a single high mass (60--80 M$_{\odot}$) O star, the specific mass of which is refined by \citet{2005A&A...430..481H} to be 56 M$_\odot$. Outflows have been extensively investigated theoretically and modeled by various authors and is usually referred to as a `champagne flow'. The model of \citet[][Figure 8]{2001ApJ...549..979K} reproduces this general structure well in circumstances where the UC~\HII\ region is near the boundaries of its molecular core and clump respectively such that it breaks out in a preferred direction. However, this interpretation does not appear to agree with the available imagery for CC. The putative champagne flow, as observed in all IRAC bands and the IRS spectral map, appears to be mis-aligned with the opening in the CC shell\footnote{The opening is clearly visible in the 3.6 cm radio imagery but it was not possible to create contour levels for Figure~\ref{fig1} to show this. (Data available from http://ecademy.agnesscott.edu/$\sim$cdepree/extreme/data.html).} by approximately 45$^{\circ}$, as the CC shell opening is in the south east rather than east.  

In Figure~\ref{fig:spec} the integrated spectrum of CC is shown, along with the integrated spectra of the core region and the eastern region. All three spectra show strong [Ne~{\sc ii}] 12.8 \micron\ emission and relatively strong silicate absorption at 9.8 \micron. The core CC region shows the lowest silicate absorption, superficially consistent with the interpretation of CC being close to the boundaries of the molecular core and clump. However, the spectrum of the eastern part shows higher extinction, which is only consistent with the champagne flow model if we assume that all the silicate absorption is in front of the flow from region CC. Given this, and the mis-alignment mentioned earlier, it seems unlikely that the eastern part of CC represents an outflow. 

\paragraph{DD} Region DD is a shell-like UC~\HII\ region with symmetric gaps in the shell. DD is referred to as `filamentary' by \citet{1997ApJ...482..307D}, but those authors mention that it could be thought of as a broken shell. In this sense it is similar to CC, however, DD is slightly larger (R = 0.43 pc) and does not have associated extended emission in the radio accompanying the emission from the central regions. In a ground based photometric MIR study of W49A by \citet{2000ApJ...540..316S}, it was shown that DD possesses a southern counterpart, `DD south', which does not correspond to any strong radio emission but did show weak [Ne~{\sc ii}] 12.8 \micron\ emission, characteristic of ionized material. DD south is coincident with a spatially small knot of weak emission in the IRAC 8 \micron\ imagery. However in the 12.8 \micron\ image presented by \citet{2000ApJ...540..316S} it is of approximately equal brightness with the main part of DD. DD south does not appear in the \textit{Spitzer}/IRS field of view, yet it is of interest as a possible counterpart to CC east. 

The integrated spectrum of region DD is shown in green in Figure~\ref{fig:spec}. It possesses strong silicate absorption ($\tau_{9.8}$ = 3.23).

\paragraph{EE} Source EE is much smaller than CC and DD, with a radius of 0.26 pc \citep{1997ApJ...482..307D}. Although detected by \citet{2000ApJ...540..316S} in ground-based mid-IR observations, very little could be derived from the observations except for a lack of emission in the [Ne~{\sc ii}] 12.8 \micron\ line which was evident in their derived SED. This contrasts with the integrated IRS spectrum of source EE (Figure~\ref{fig:spec}), which exhibits clear [Ne~{\sc ii}] emission. This effect likely derives from the fact that the [Ne~{\sc ii}] emission observed in the IRS map is weak compared to the continuum (as is obvious in Figure~\ref{fig:spec}). 

\paragraph{GG}The radio source GG is the smallest in the field of view (R = 0.022 pc; \citealt{1997ApJ...482..307D}). Radio observations revealed it to be consistent with a B0 spectral type exciting star, the only labeled region not thought to be generated by an O star. 

As one might expect given the later spectral type of the exciting star, the integrated spectrum of region GG contains the weakest ionized lines of any of the UC~\HII\ regions. In every other respect though, the spectrum of region GG is very similar to that of the other regions, including the presence of strong silicate absorption.

\paragraph{II} Radio source II lies just to the northeast of the Welch ring. In the radio, it appears as a large, faint, shell of emission with a radius of 0.65 pc \citep{1997ApJ...482..307D}. It is the only complete shell present in the \textit{Spitzer}/IRS field of view, although it is difficult to make out the shell in the \textit{Spitzer}/IRS data as roughly half of it is superimposed upon emission from the arc region. In the IRAC 8 \micron\ imagery, II appears as a void of emission which is perfectly aligned with the shell structure seen in the radio. This is possibly explained by the integrated spectrum of region II being affected by the deepest silicate absorption of all of the UC~\HII\ regions in the sample with an optical depth of $\tau_{9.8}$ = 4.17.

\subsection{Maps}\label{sec:maps}
In the following Section, maps of the measured features are shown, described in detail and compared in terms of their appearance. For ease of navigation, this section is subdivided with some general comments about all the maps to begin, followed by discussion of the extinction map presented earlier in Figure~\ref{fig:ext}. We then discuss the PAH emission maps (Section~\ref{sec:pahmaps}, Figures~\ref{fig:slmapsdr} \& \ref{fig:slextras}); ionized line emission (Section~\ref{sec:ionmaps}, Figure~\ref{fig:slionmaps}) and the continuum emission (Section~\ref{sec:contmaps}, Figure~\ref{fig:slcont}). The maps are presented in their dereddened form where the extinction correction was applied as described in Section~\ref{sec:ext}.

\subsubsection{General Comments}\label{sec:mapsgc}
In general, the maps of emission features split into two distinct groups, mainly based on the appearances of CC and DD: those which peak on the shells of the \HII\ regions and those which peak towards the exciting star of the UC~\HII\ regions. While there are some deviations from this picture in specific maps, these are relatively minor and will be discussed in specific subsections.

The set of maps which peak on the shells of the UC~\HII\ regions includes all of the PAH bands and their underlying plateaus. The group of maps peaking at the cores of the UC~\HII\ regions contains the continua and all of the ionized lines except [Ne~{\sc ii}] 12.8 \micron. The [Ne~{\sc ii}] line at 12.8 \micron\ (Figure~\ref{fig:slionmaps}), behaves in an intermediate way: its structure for CC and DD appears to be a fairly uniform emission across each region.

\subsubsection{Extinction Map}
The morphology of the silicate absorption shown in Figure~\ref{fig:ext} is interesting in that it is very different from the morphologies evident in each IRAC band as well as the maps of PAH and atomic line emission presented elsewhere in this section. Comparing the map with the spectra in Figure~\ref{fig:spec} we see the core region of CC possesses the weakest silicate feature and this corresponds nicely with a minimum in the map. Conversely, the deepest silicate features occurred for UC~\HII\ region II, which in the map appears to be associated with some of the highest degrees of silicate absorption. Along with the smoothness of the resulting map, the consistency supports the idea that this discrepant morphology is not an artificial effect. A somewhat similar situation was encountered for IRAS 12063-6259 by \citet{2013ApJ...771...72S}, where the pattern of the silicate absorption did not match that of other MIR emission components. There are several potential physical geometries which could explain this phenomenon. In both examples (W49A and IRAS 12063-6259) the sources are located at distances of 10 kpc or greater along sightlines passing through copious galactic plane material, implying that the strange distribution of silicate absorption could reflect the spatial layout of molecular clouds along the line of sight. However, this is unlikely to be the full explanation -- upon inspection of Figure~\ref{fig:ext} the patches of the highest extinction are clustered around the UC \HII\ regions near the core of W49A with little strong extinction around the periphery. This argues in favour of the extinction being a composite of that along the line of sight and within W49A. The discrepant pattern of extinction might then be due to a host of processes within - anything from dust destruction around the massive stars to the collective effect of the massive star winds changing the spatial distribution.

\citet{2005A&A...430..481H} deduce that the average total foreground galactic line of sight extinction is around $A_K = 2.1$. This was derived by fitting an extinguished luminosity function to the observed color-color diagram of the W49A stars detected in a 5\arcmin $\times$ 5\arcmin\ field, so this value should be treated as a spatial average, especially in light of the large variations we detect in a much smaller field. In a study of the CH$^+$ emission towards W49A \citep{2012A&A...540A..87G}, it was found that the total number of hydrogen atoms along the W49A line of sight is $460 \times 10^{20}\textrm{ cm}^{-2}$ $\pm$ $50\%$. This value corresponds to an average A$_K$ of around 2.3, which is in good agreement with the result of \citet{2005A&A...430..481H}.

Both previous studies were measuring a slightly different quantity than that found with the adapted Spoon method because of their methodologies: the average foreground extinction. In contrast the Spoon method measures both the foreground extinction detected by the other methods and extinction generated within W49A. For \citet{2005A&A...430..481H}, this spatial average arises from the fitting process of one luminosity function to the color color plot, while for \citet{2012A&A...540A..87G} it arises from a lack of spatial resolution. Therefore, the range of $A_K$ derived using the iterative Spoon method agrees with the quoted figures even though there are areas of higher and lower extinction. 

\subsubsection{PAH Emission Maps}\label{sec:pahmaps}

\begin{figure*}
  \centering
 
  \resizebox{\hsize}{!}{%
    \includegraphics{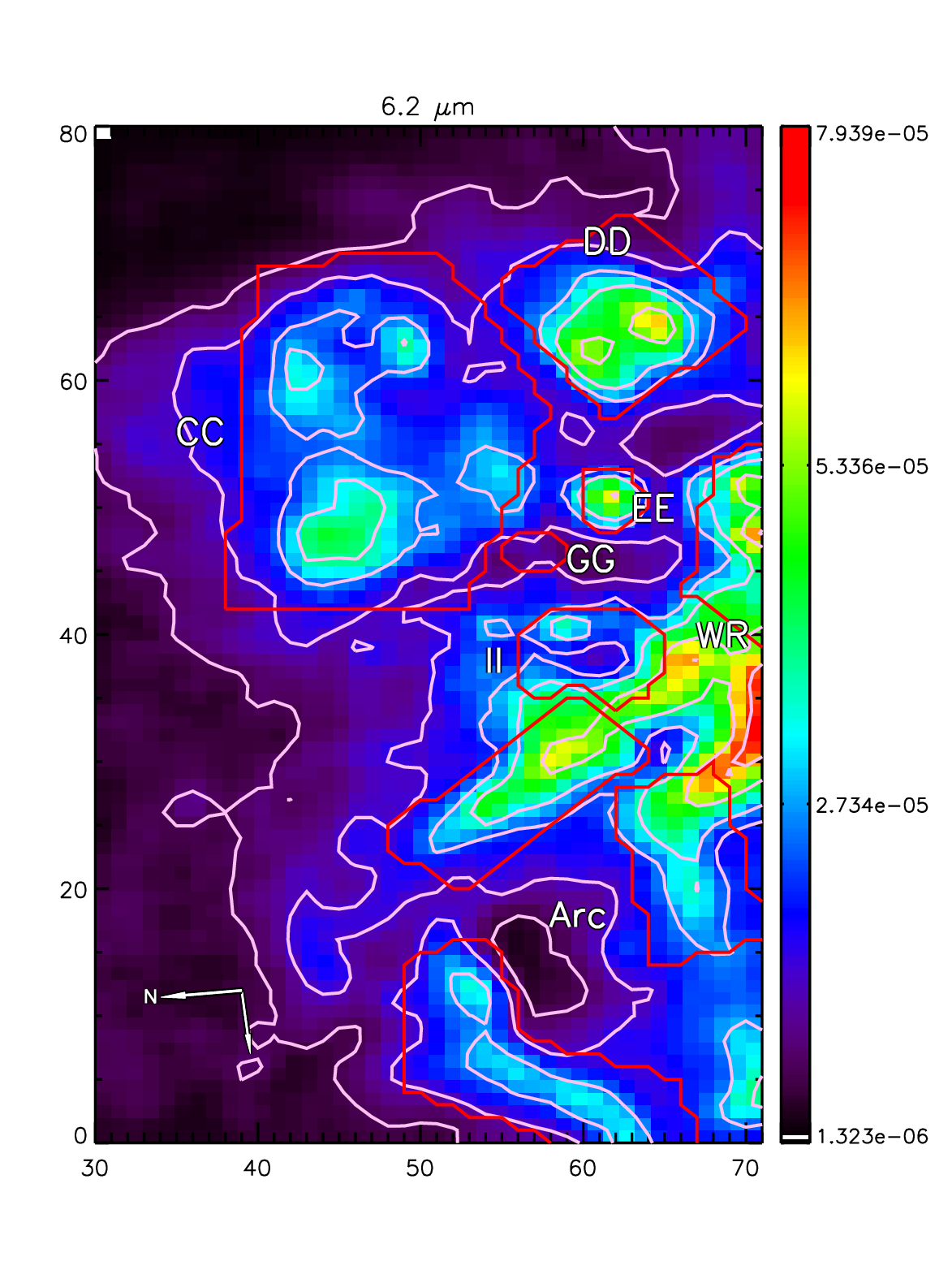}
    \includegraphics{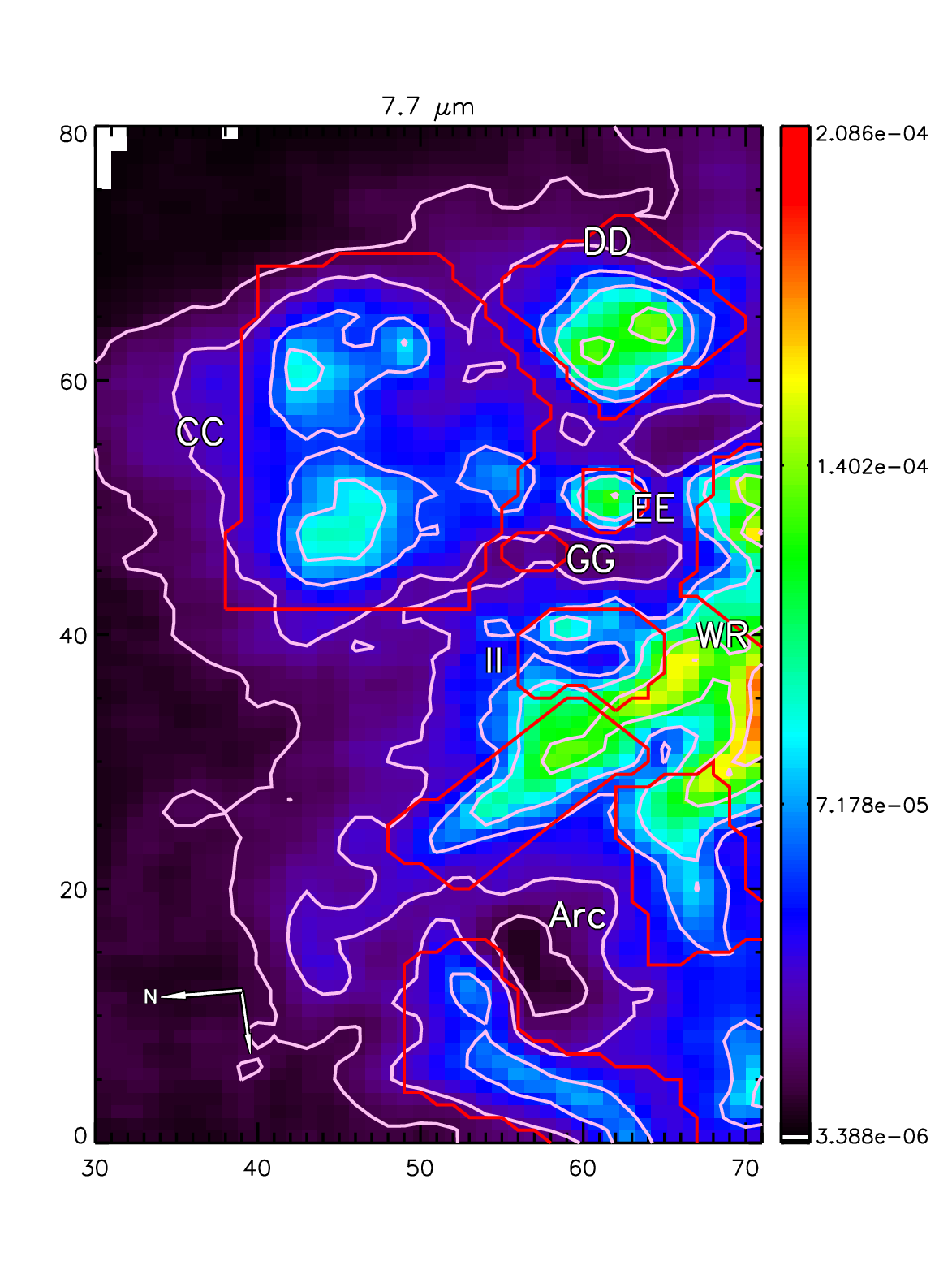}

}
  \resizebox{\hsize}{!}{%
    \includegraphics{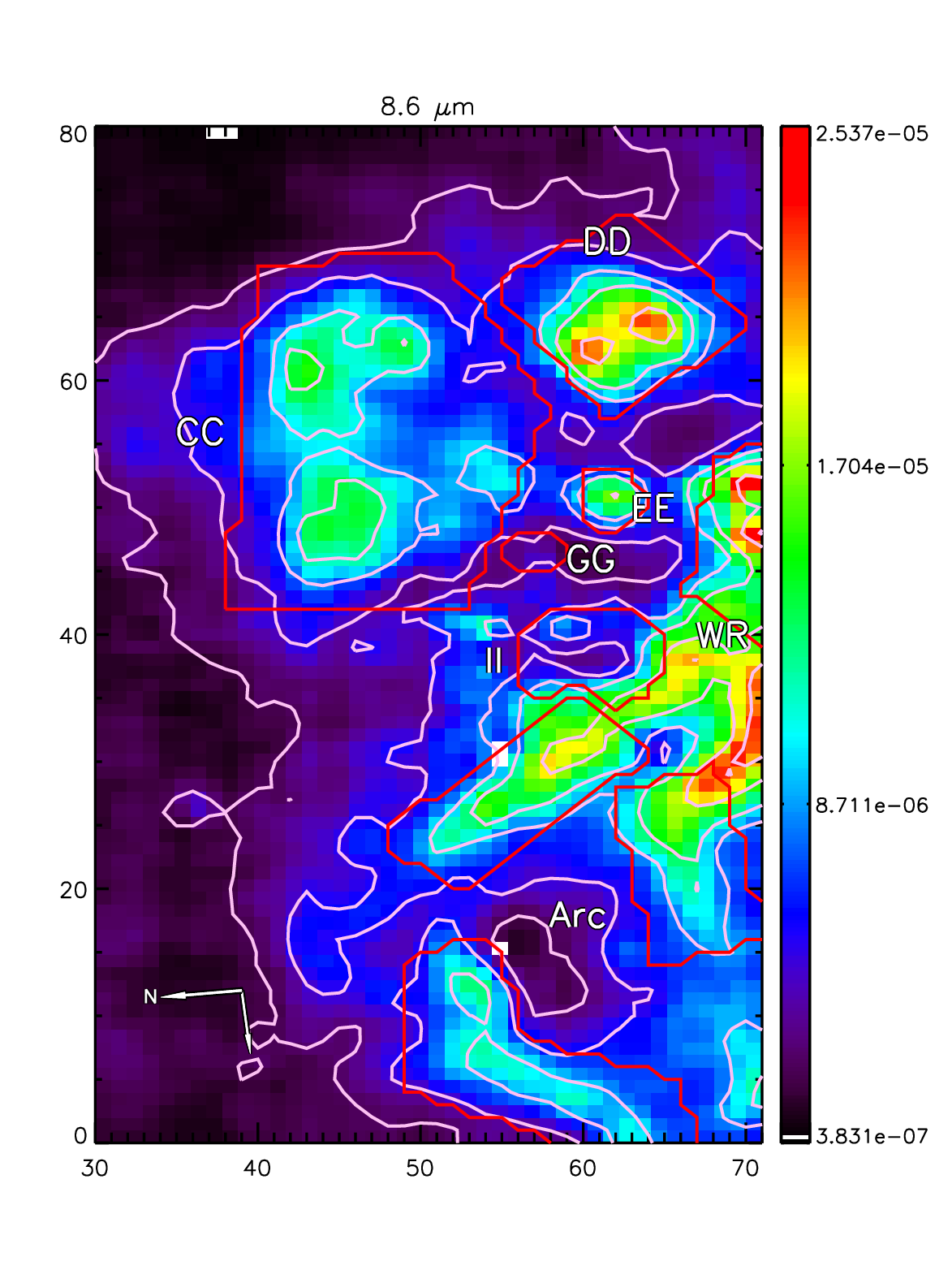}
    \includegraphics{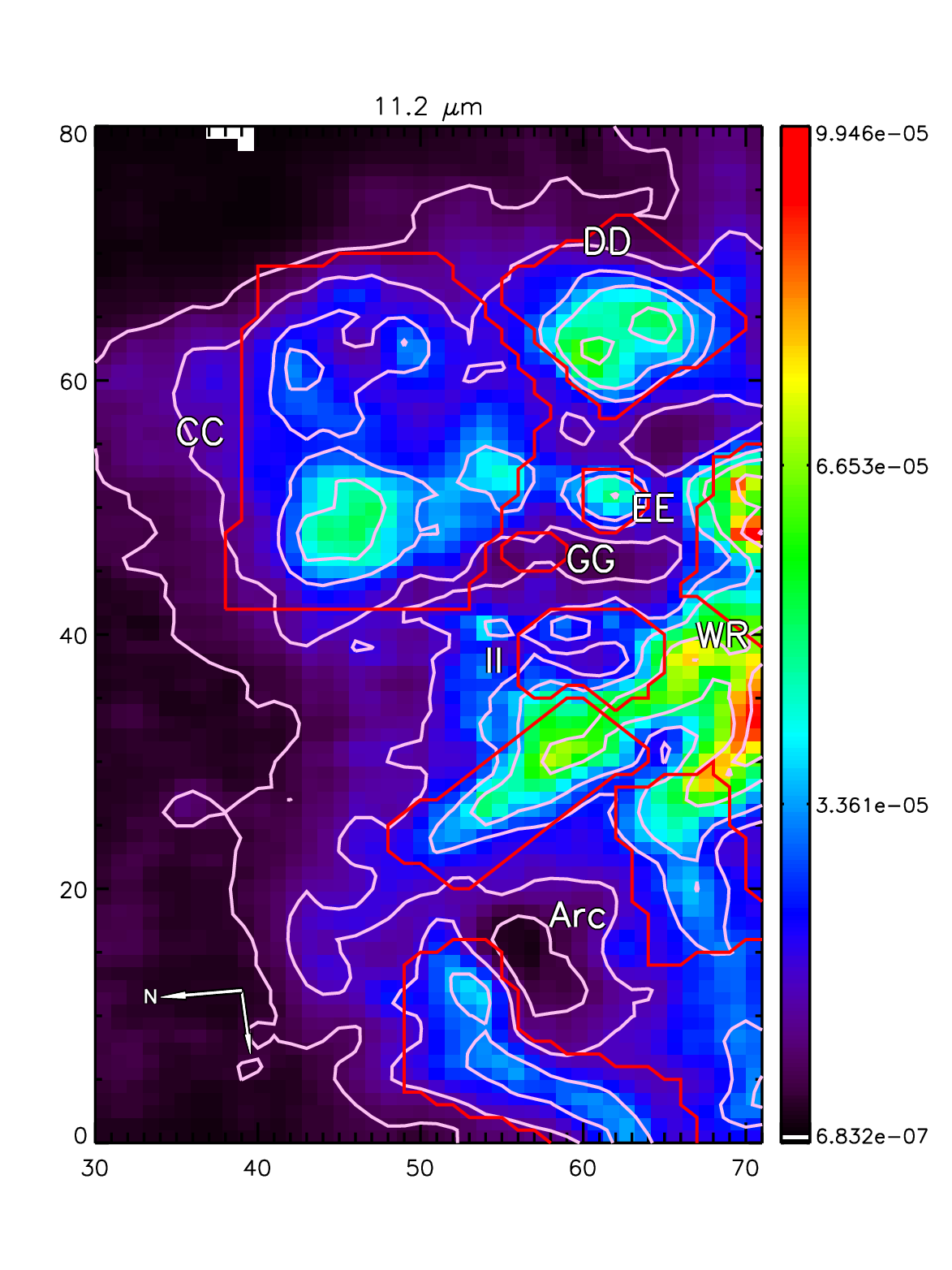}
}
    \caption{Dereddened maps of the integrated intensity of the PAH emission bands across the observed part of W49A in units of W m$^{-2}$ sr$^{-1}$. Clockwise from top left, 6.2 \micron, 7.7 \micron, 11.2 \micron\ and 8.6 \micron. In each panel the pink and red contours represent the 7.7 \micron\ emission and the regions selected for each UC~\HII\ region and arc section respectively. Data which fell below a S/N cut of three was excluded and these regions are represented as white pixels.}
    \label{fig:slmapsdr}
\end{figure*}

In Figure~\ref{fig:slmapsdr}, maps of the dereddened strengths of the main PAH emission bands are shown. Overall the structures seen in the PAH emission maps are very similar to the structure of the IRAC 8 \micron\ image shown in Figure~\ref{fig1}. 

The main characteristics of the 7.7 \micron\ PAH emission map are as follows: strong emission tracing the shells of the CC/DD UC~\HII\ regions; UC~\HII\ region EE appearing as a point source; UC~\HII\ region II displays some traces of emission around its perimeter, with a very bright shell section to the south east; the northern section of the Welch ring is very patchy and the emission does not seem to follow the distribution of UC~\HII\ regions. The arc region is very smooth and well defined, but much fainter than the UC~\HII\ regions. Additionally, the shell of CC is not traced in its entirety, there is a gap at the south east section which corresponds to the gap visible in the 3.6 cm radio imagery. Finally, the brightest area of emission in the map occurs to the east of the Welch ring and is not associated with any detected continuum source. 

The 6.2 \micron\ emission closely follows the morphology of the 7.7 \micron\ emission, with two small exceptions in the shells of CC and DD: firstly, the south side of CC is approximately the same intensity as the north for the 6.2 band, while for 7.7 the south side is weaker; secondly, the south west side of DD has a higher intensity than the north east side for 6.2, while for 7.7 the two sides are of similar intensity. These two components are approximately the same brightness in the 7.7 and 8.6 \micron\ maps. Indeed, the 8.6 \micron\ emission seems to trace the 7.7 emission very closely in the ionized regions. The discord between 6.2 and 7.7 is unexpected, as usually these bands are found to vary together \citep{2008ApJ...679..310G}. Traces of this phenomenon are seen again in the correlation plots in Section~\ref{sec:corrs} and discussed in Section~\ref{sec:struct}.

The 11.2 \micron\ emission map appears somewhat different from the 7.7 \micron\ map. The CC and DD shells appear similar to those seen at 7.7 \micron, albeit with a deeper central minimum which for CC extends towards the south east through the gap in the shell. However, the relative brightness has changed significantly with the CC shell becoming much fainter than DD. In the 11.2 \micron\ emission map, the eastern section of UC~\HII\ region CC is actually brighter than the core/western section. 

In each of the four maps mentioned thus far, there is a seemingly continuous structure starting from the Welch ring and connecting to the CC east emission via EE. 

\begin{figure*}
  \centering
  \resizebox{\hsize}{!}{%
    \includegraphics{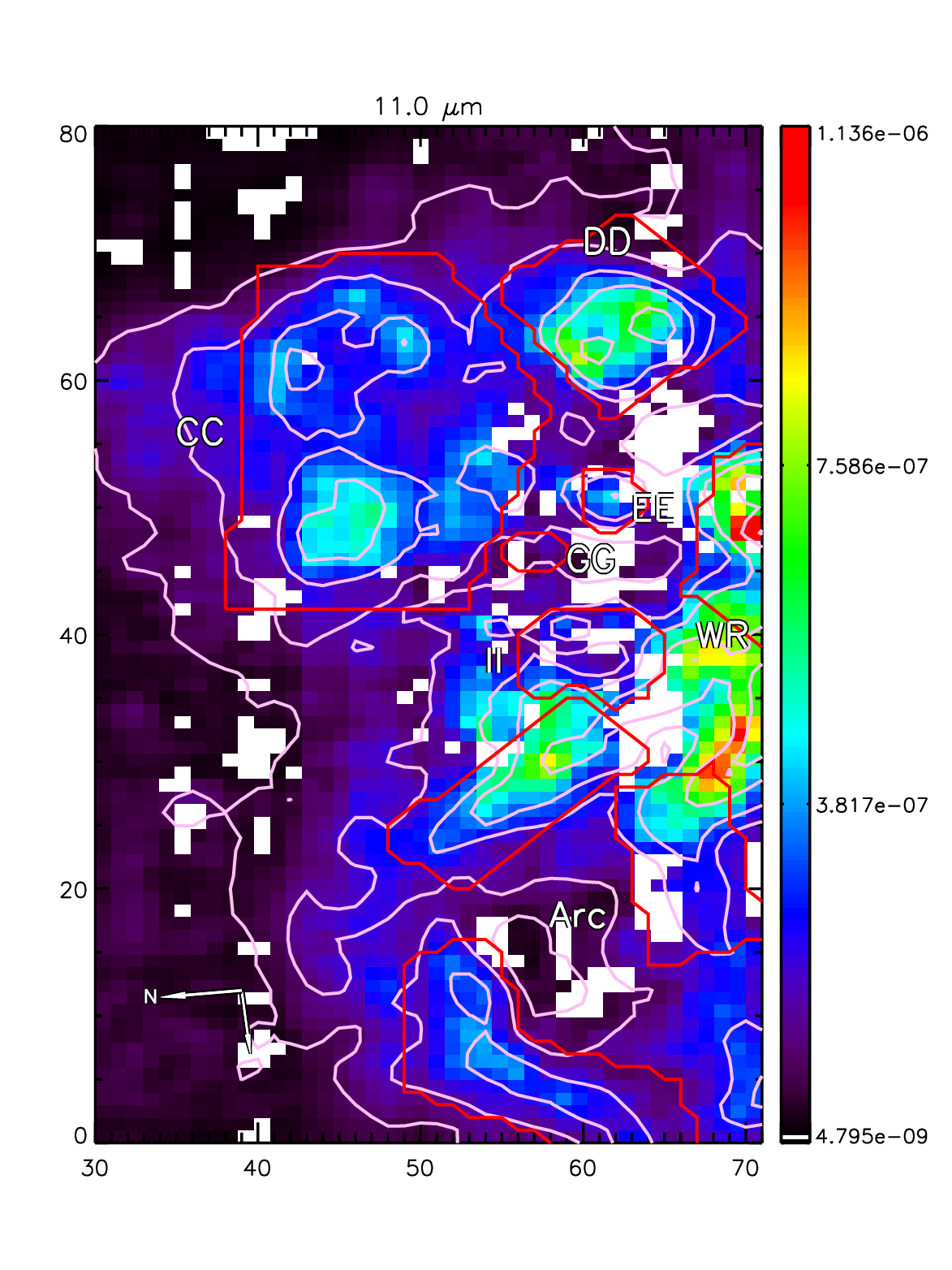}
    \includegraphics{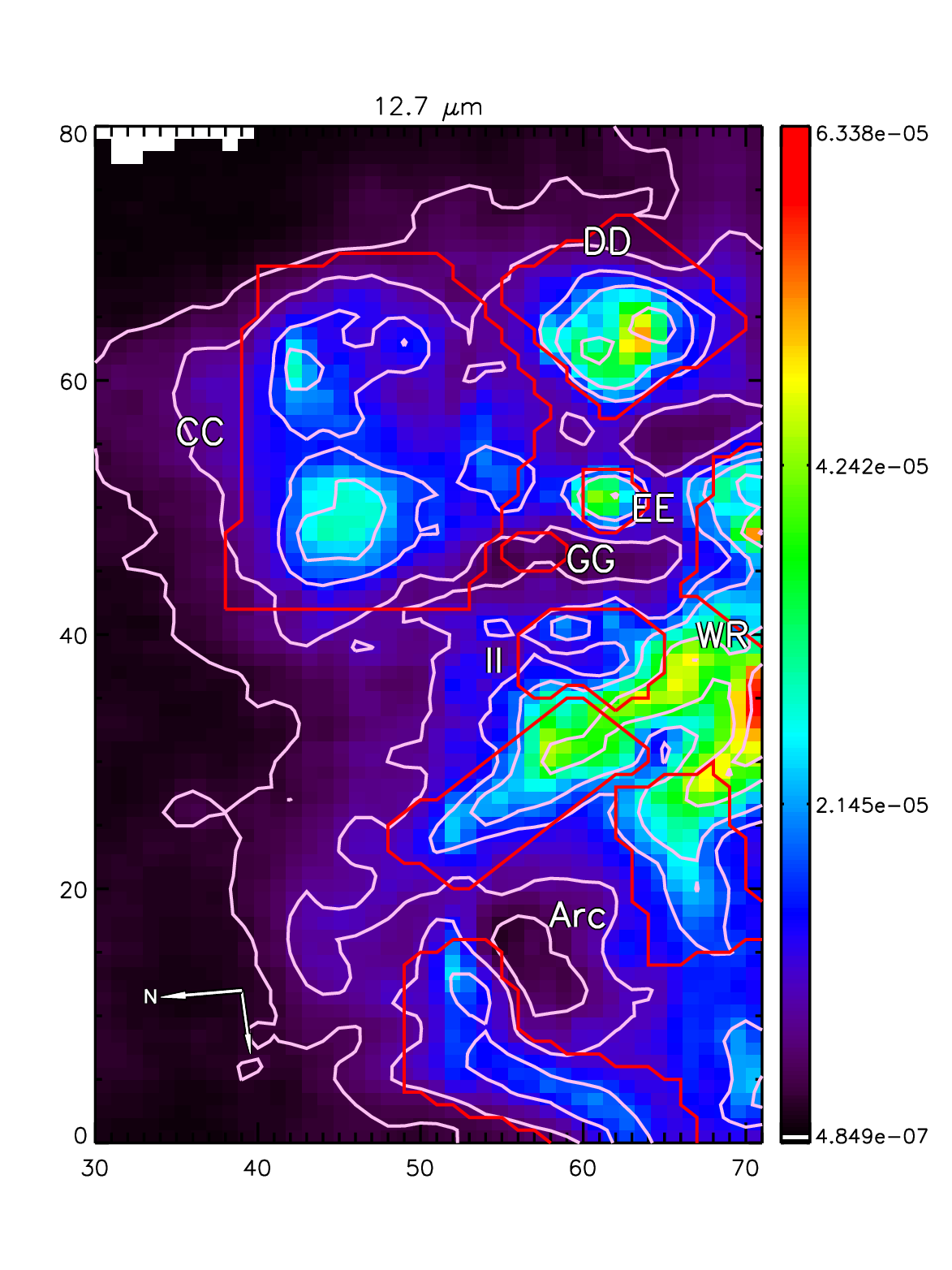}
}
  \resizebox{\hsize}{!}{%
    \includegraphics{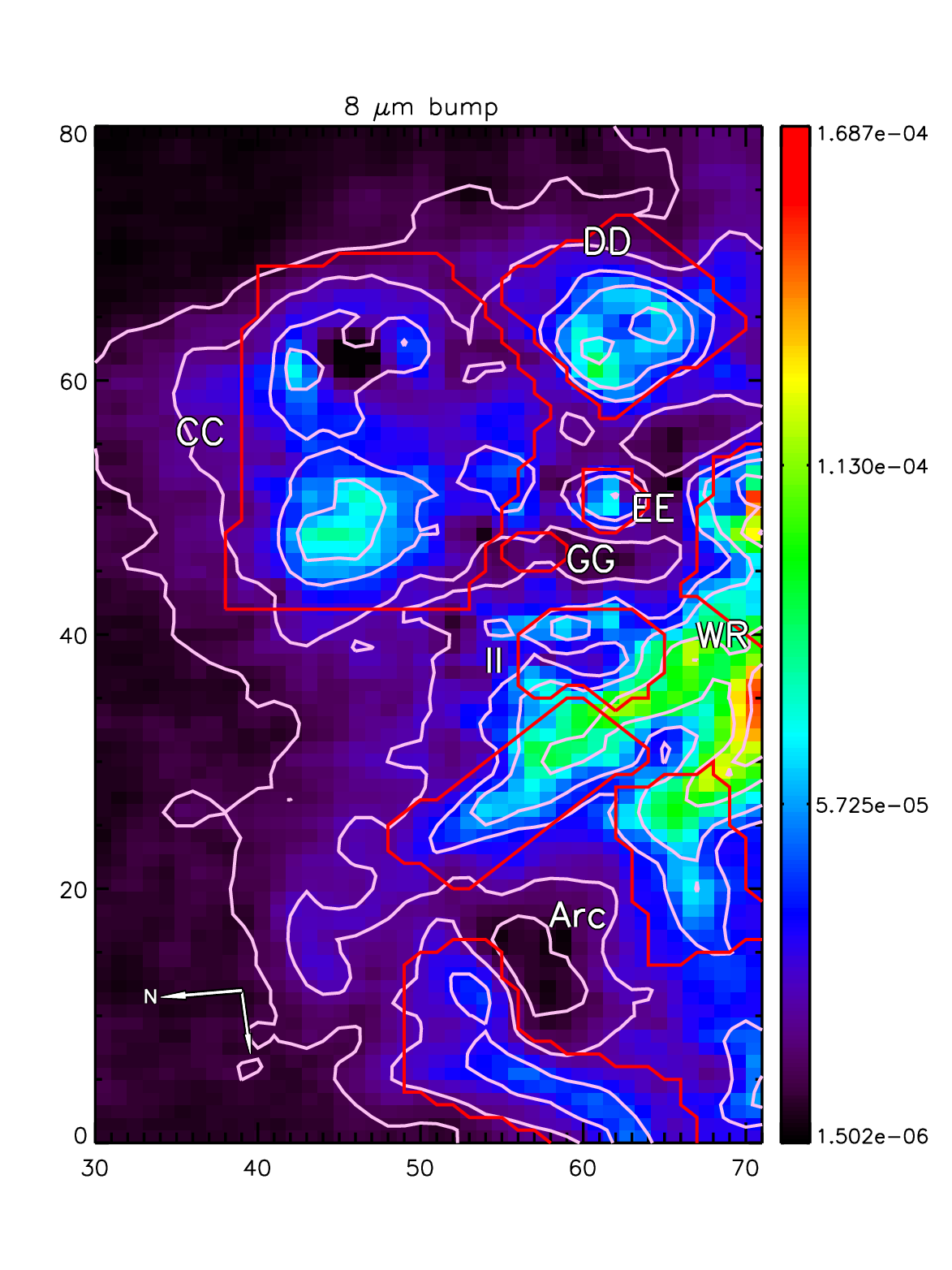}
    \includegraphics{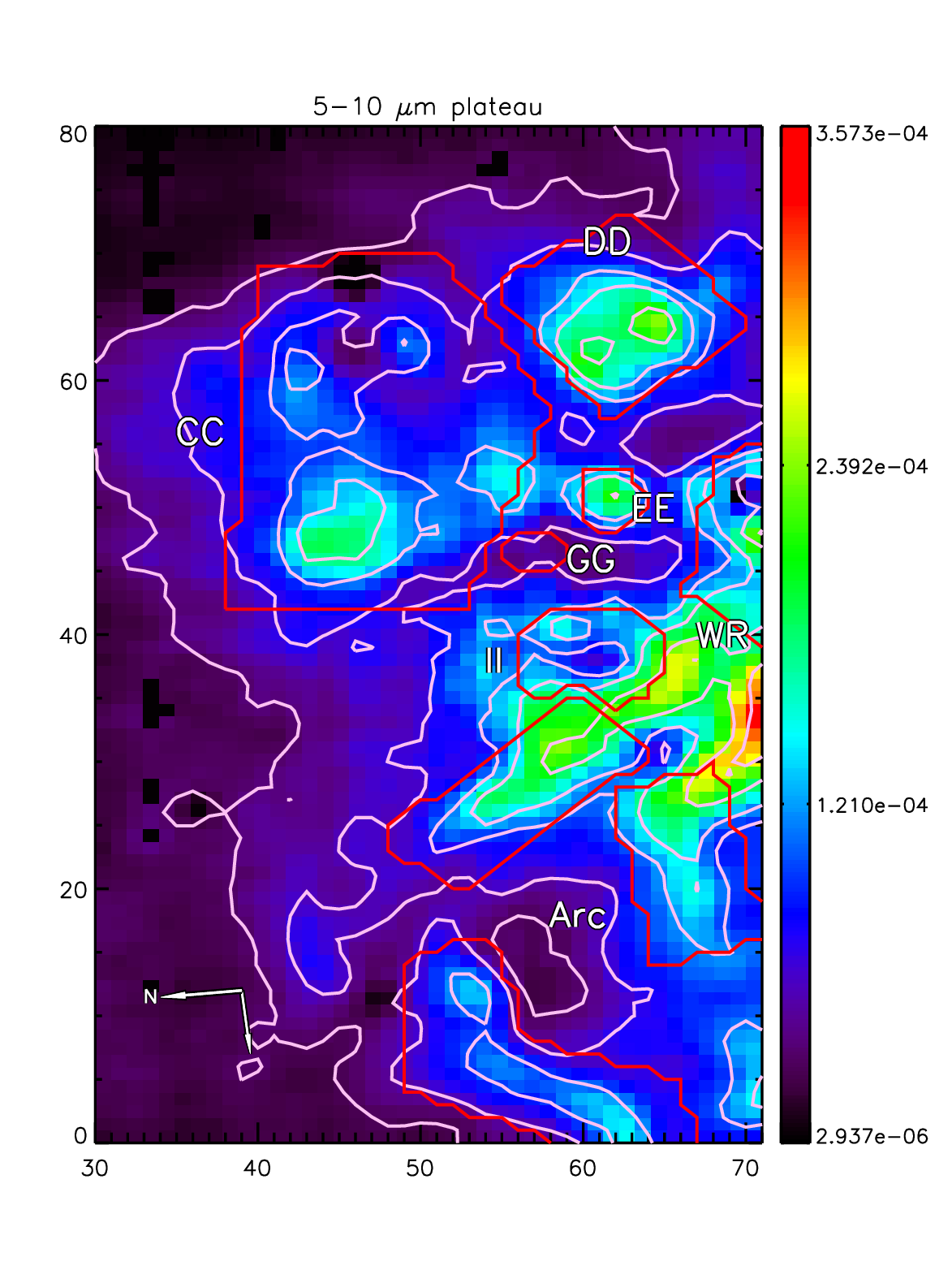}
}
    \caption{Maps of the integrated intensity of the weaker PAH features and plateaus in units of W m$^{-2}$ sr$^{-1}$, clockwise from top left: 11.0 \micron; 12.7 \micron; 5-10 \micron\ plateau and the 8 \micron\ bump. Contours, labels and S/N cut as in Figure~\ref{fig:slmapsdr}.}
    \label{fig:slextras}
\end{figure*}

\smallskip
In Figure~\ref{fig:slextras}, two of the weaker PAH emission bands, at 11.0 and 12.7 \micron, are shown. The 11.0 \micron\ band map seems to follow the other PAH emission maps in peaking around the perimeter of UC~\HII\ region CC, albeit at a slightly greater radius than the other bands. This result should be treated with caution though as the 11.0 \micron\ band has very low S/N and some systematic problems. 

The map of 12.7 \micron\ PAH emission is more intriguing, displaying a mixture of features seen in previous maps. Most striking are the CC/DD shells as both appear much brighter on one side (as in the 6--9 \micron\ PAH feature maps) however this is combined with the lower emission levels of the 11.2 \micron\ feature map in the center of CC. In the case of CC this occurs on the north side while for DD the south is brighter. The eastern region of CC is also very bright and is more compact than seen in previous bands, with less  emission extending along the structure connecting CC, EE and the Welch ring.

We have defined two extended broad emission features in the 5--10 \micron\ region (see Section~\ref{sec:dra}, Figure~\ref{fig:decomp}), the 5--10 \micron\ plateau and the 8 \micron\ bump. Maps of the emission of these components are shown in Figure~\ref{fig:slextras}. Both of the plateau maps share the unusual properties of having deep minimas in the center of the CC and II shells. In the case of DD, both trace the shell, but the 8 \micron\ bump has much lower emission in the center while for the 5--10 \micron\ plateau there is constant emission across the center. In each map the CC shell is broken and fainter than the DD shell. The II shell is somewhat traced in both maps, but stronger in the 5--10 \micron\ map.  The 5--10 \micron\ map shows the connection between CC east, EE and the Welch ring even more clearly than the 11.2 \micron\ map. Additionally, the shape of the arc filaments differs between the two, with the 5--10 \micron\ map strongly tracing the filament near UC~\HII\ region II and this emission being weaker in the 8 \micron\ bump map. The map of the longer wavelength continuum is much noisier than the previous two, but appears to follow the 8 \micron\ bump rather than the 10--15 \micron\ plateau. 

\subsubsection{Emission Line Maps}\label{sec:ionmaps}
\begin{figure*}
  \centering
  \resizebox{\hsize}{!}{%
    \includegraphics{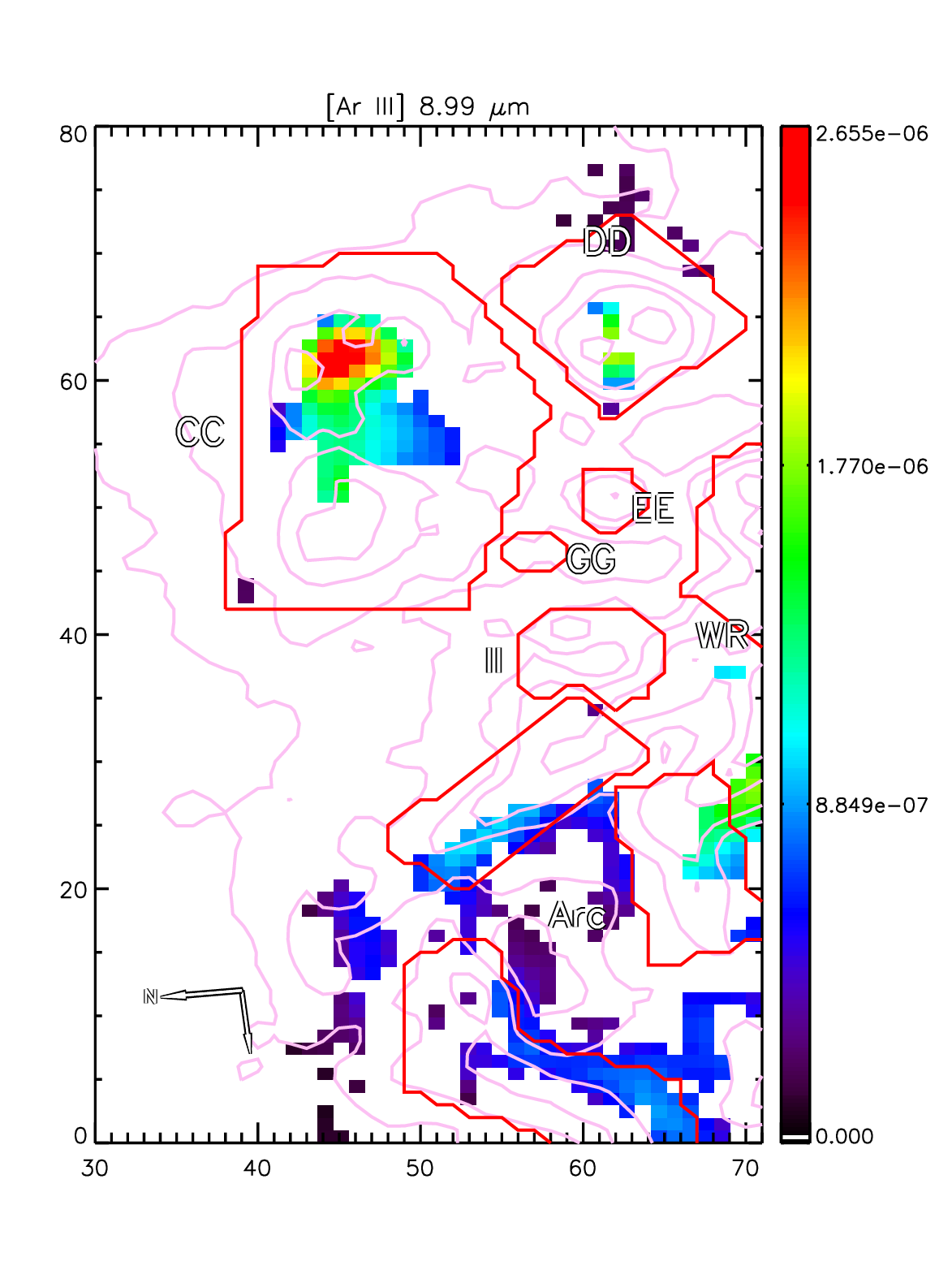}
    \includegraphics{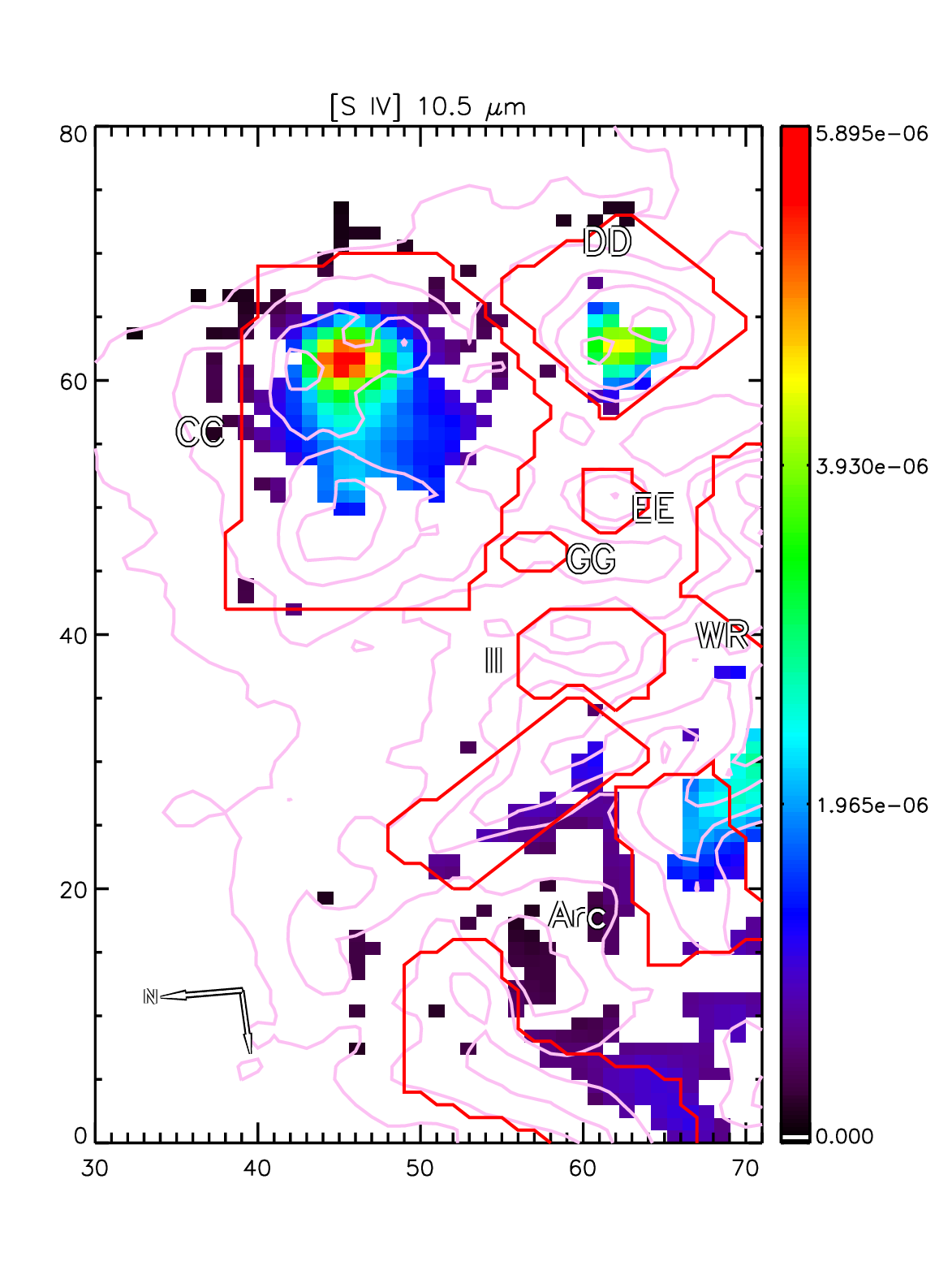}
}
  \resizebox{\hsize}{!}{%
    \includegraphics{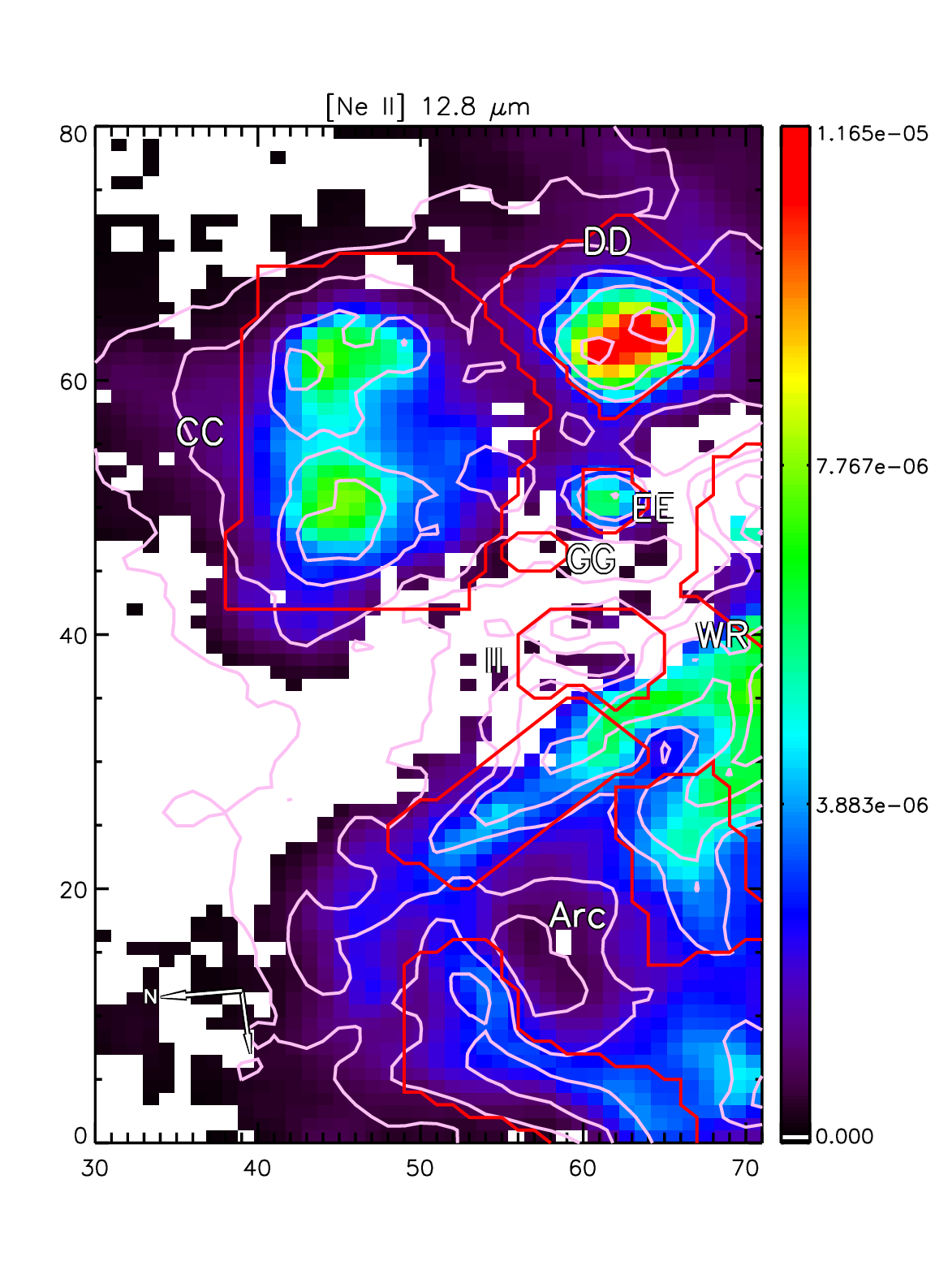}
    \includegraphics{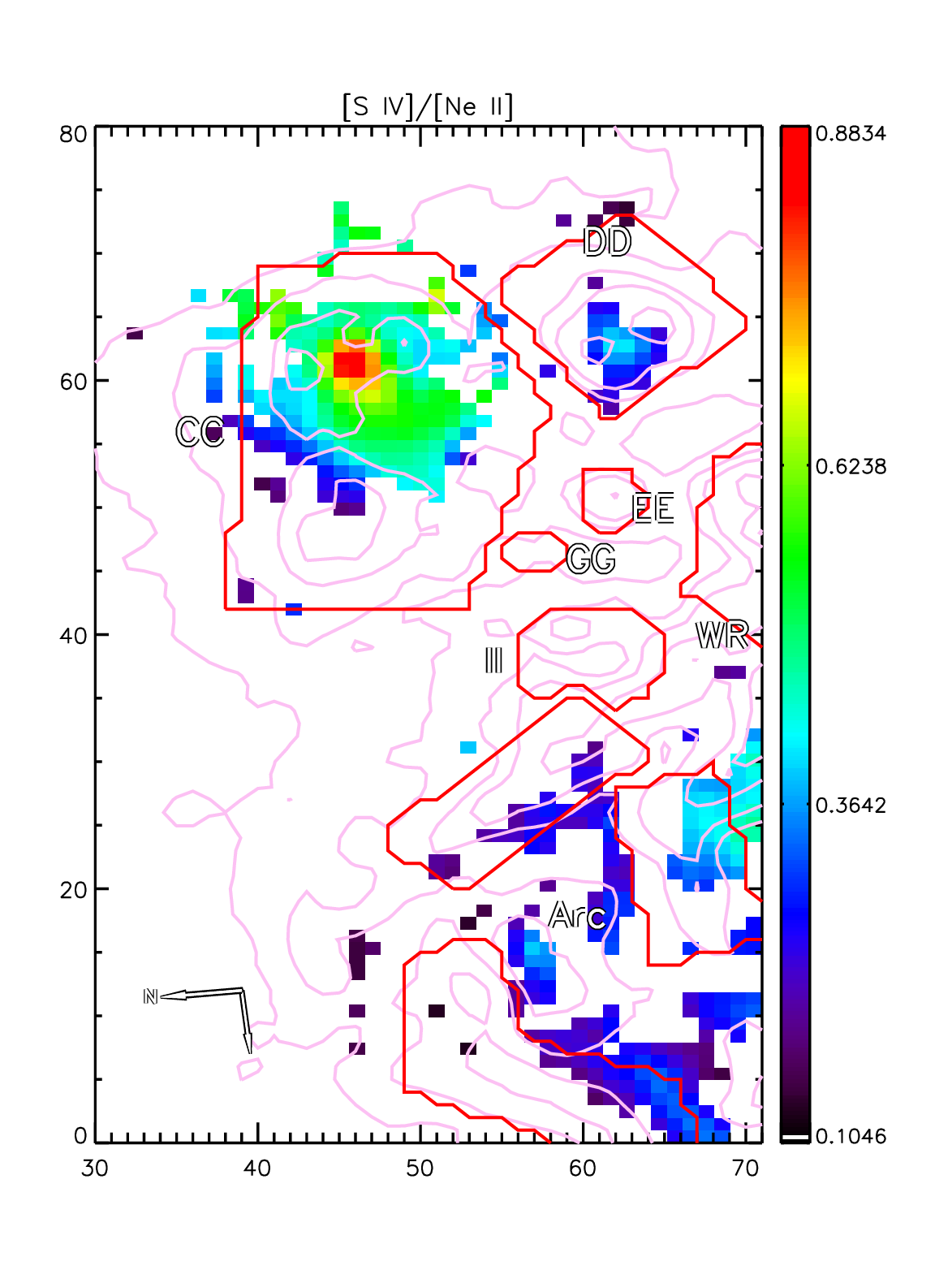}
}

    \caption{Maps of ionized line emission and line ratios across the observed part of W49A in units of W m$^{-2}$ sr$^{-1}$. Clockwise from top left: [Ar~{\sc iii}] 8.99 \micron; [S~{\sc iv}] 10.5 \micron; [S~{\sc iv}] 10.5 \micron/ [Ne~{\sc ii}] 12.8 \micron; [Ne~{\sc ii}] 12.8 \micron. Contours and labels as in Figure~\ref{fig:slmapsdr}, S/N cut reduced to two for ratio map. }
    \label{fig:slionmaps}
\end{figure*}

In Figure~\ref{fig:slionmaps}, the emission from ionized lines in the W49A field of view is shown. The [Ne~{\sc ii}] 12.8 \micron\ line requires the lowest ionization energy of all the lines detected (IP = 21.56 eV). Consequently, the [Ne~{\sc ii}] map displays a more uniform appearance in the UC~\HII\ regions CC and DD. In addition, the UC~\HII\ region EE is a bright point source and emission from II falls below the S/N cutoff. 

The higher ionization potential lines ([S~{\sc iv}], [Ar~{\sc iii}]; IPs = 34.83, 27.63 eV resp.), show much weaker emission than [Ne~{\sc ii}] thus large sections of each map do not meet the imposed S/N $>$ 3 cutoff. However, there are some subtle differences between the [S~{\sc iv}] 10.5 \micron\ and [Ar~{\sc iii}] 8.99 \micron\ maps. The [S~{\sc iv}] map peaks more sharply towards the central sources of the UC~\HII\ regions. The [Ar~{\sc iii}] map also shows some emission around the arc region, while [S~{\sc iv}] has very low emission there. The [Ar~{\sc iii}] emission occurs at the inner edge of the arc regions.
  
The line ratio of [Ne~{\sc iii}] 15.6 \micron\ / [Ne~{\sc ii}] 12.8 \micron\ is usually employed as a proxy for ionization. The \textit{Spitzer}/IRS spectra of W49A do not extend to 15.6 \micron\ so a different tracer must be used. It has been shown that the [S~{\sc iv}] 10.5 \micron\ / [Ne~{\sc ii}] 12.8 \micron\ line ratio mimics the [Ne~{\sc iii}] / [Ne~{\sc ii}] ratio under a very wide range of conditions \citep{2008MNRAS.391L.113G}. 

Even though the S/N is low which restricts the area for which there is data, it is clear that the morphology of the line ratio map differs from those discussed previously. The CC region in particular shows a very sharp transition from low ratios to much higher ratios in terms of [S~{\sc iv}] / [Ne~{\sc ii}] on a diagonal line from south-east to north-west just to the east of the core of UC~\HII\ region CC. The [S~{\sc iv}] / [Ne~{\sc ii}] line ratio is a factor of two weaker in the core of DD than CC, implying a much lower level of ionization relative to CC. Region CC also has an odd appearance, with the lobes of PAH emission to the north and south visible in the contour maps corresponding to patches of lower ionization on either side of the central peak (e.g. at $x$=42, $y$=61).

\subsubsection{Continuum Maps}\label{sec:contmaps}

\begin{figure*}
  \centering
  \resizebox{\hsize}{!}{%
    \includegraphics{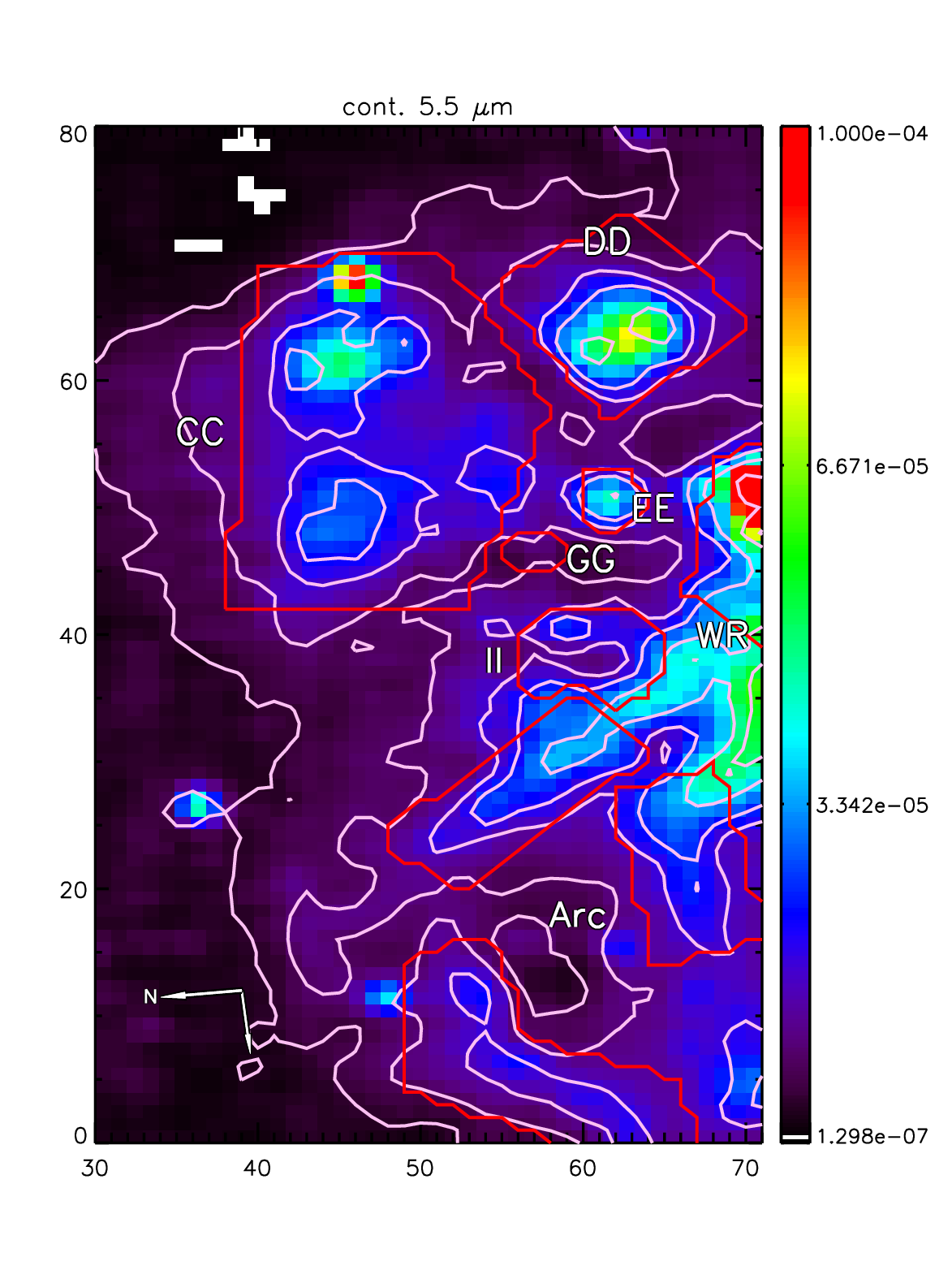}
    \includegraphics{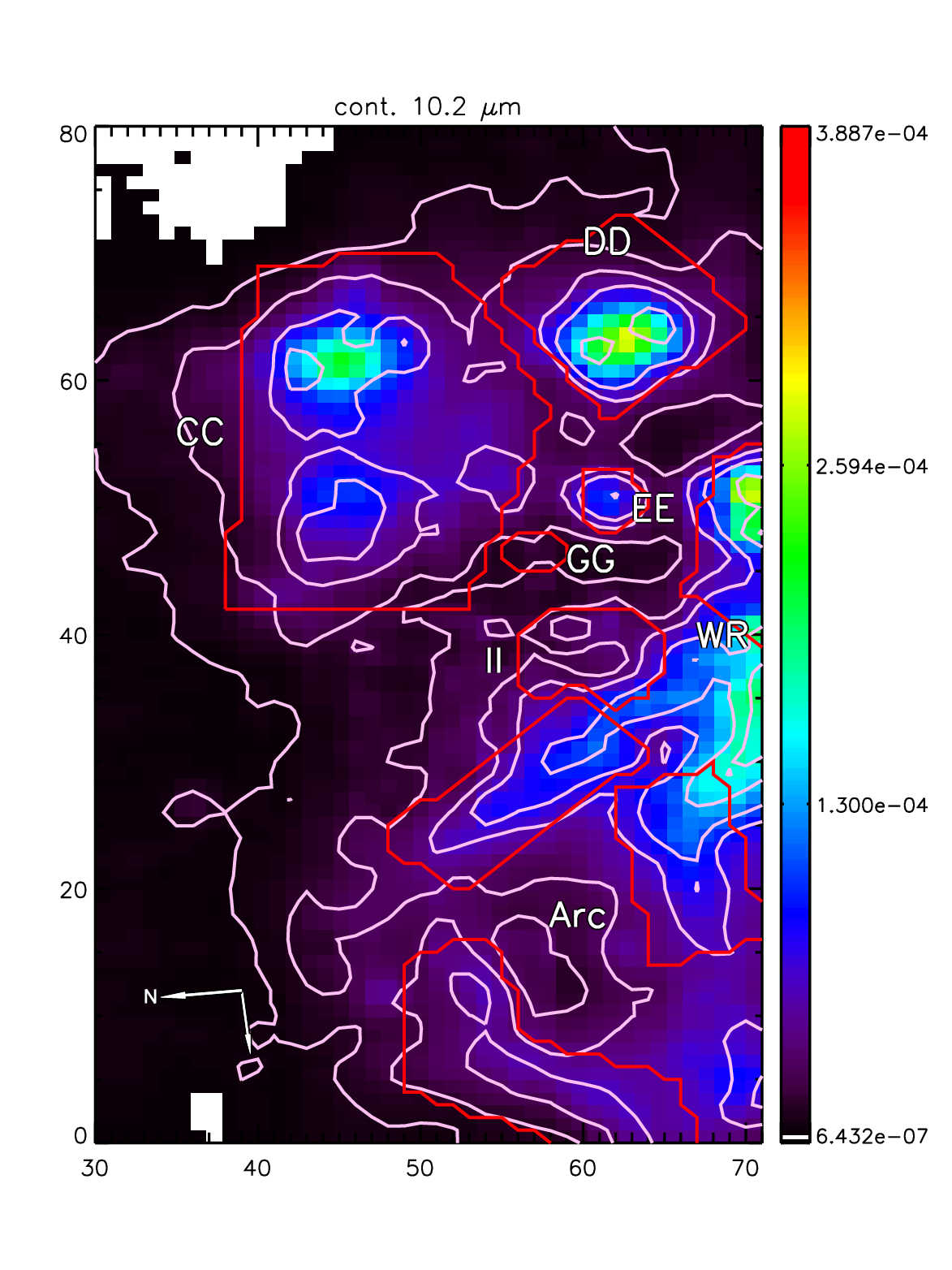}
    \includegraphics{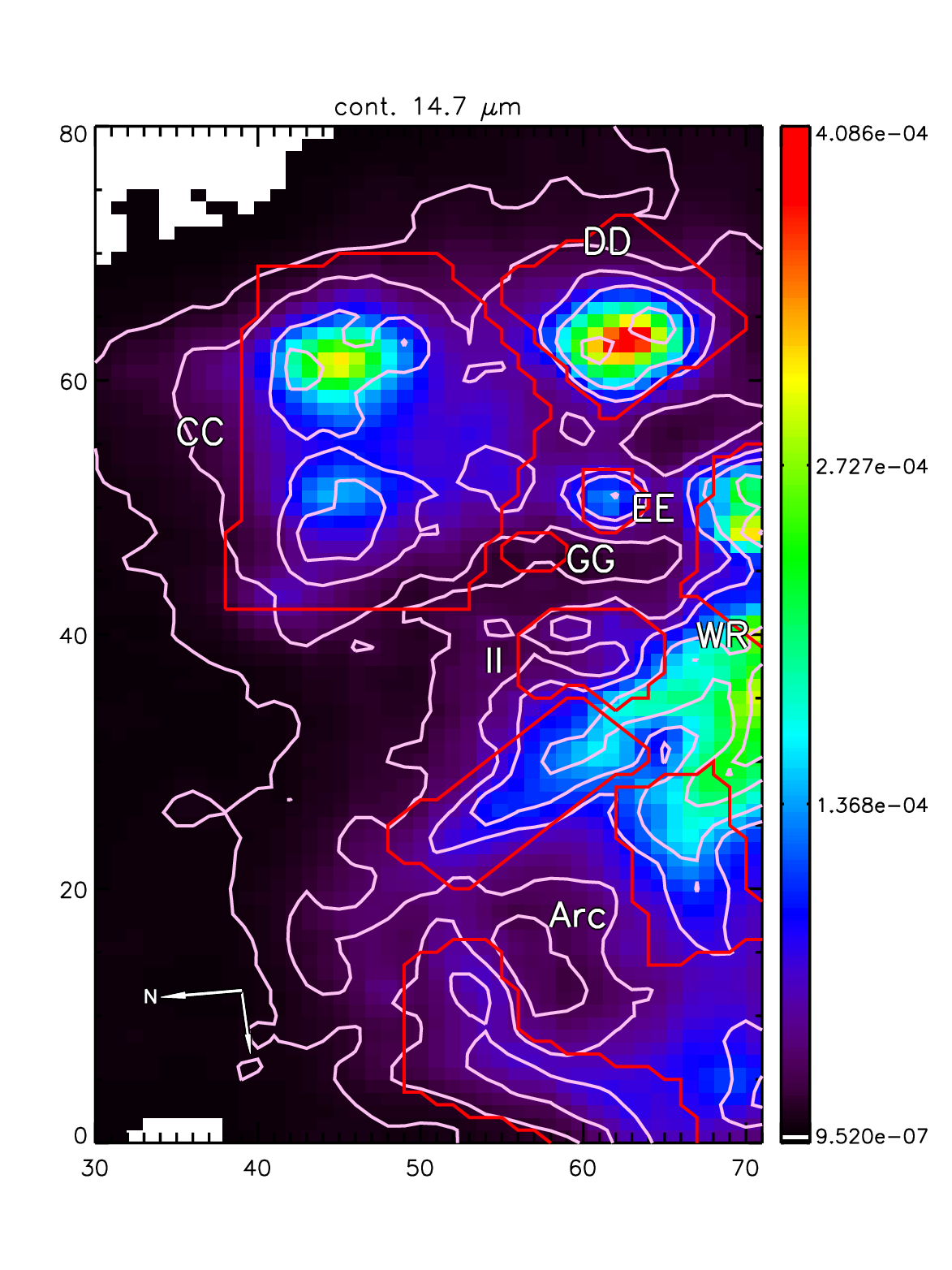}
}
    \caption{Maps of continuum emission in units of W m$^{-2}$ sr$^{-1}$ \micron$^{-1}$: \textit{left:} 5.5 \micron; \textit{centre:} 10.2 \micron; \textit{right} 14.7 \micron. The color scale of the 5.5 \micron\ map has been capped at 0.001 W m$^{-2}$ sr$^{-1}$  \micron$^{-1}$ to increase the contrast. Contours and labels as in Figure~\ref{fig:slmapsdr}.  }
    \label{fig:slcont}
\end{figure*}

The continuum emission, shown for 5.5, 10.2 and 14.7 \micron\ in Figure~\ref{fig:slcont}, traces the centers of the CC and DD UC~\HII\ regions. Some weaker features are also evident: there is a peak near EE and some weak emission connects CC east, EE and the Welch ring tracing the `bar' evident in the PAH emission maps. Otherwise, faint emission traces the strong PAH emission shown by the 7.7 \micron\ contours. There is also very faint emission from the arc regions in all continuum maps.

There are several extra sources evident in the continuum maps which do not show up in the PAH maps, line emission maps or the radio maps (for example, at coordinates [46,68] and [37,27]). These correspond to foreground stars which are evident in the Glimpse/IRAC [3.6] observations of W49A  \citep{2003PASP..115..953B,2009PASP..121..213C}.

\subsubsection{Maps Summary}

In general, there are remarkable similarities between all of the maps: each broadly trace the emission visible in the IRAC bands. The deviations which we have described are subtle details against a background of broad consistency. The main conclusions from inspection of the maps are that:
\begin{enumerate}
\item the usual inter-relationship between the strengths of the 6--9 \micron\ PAH bands subtly break down in the rims of the UC~\HII\ regions,
\item the maps of the plateau emission follow the PAH band emission, 
\item the continuum emission is strong in the centers of the UC~\HII\ regions,
\item there is stratification of emission in the arc segments, with [Ar~{\sc ii}] tracing the inner edges, followed by continuum emission and then PAH bands.
\end{enumerate}

\subsection{Feature Correlations}\label{sec:corrs}
To observe the correlations between relative PAH band strengths, the bands are usually divided by a third PAH band. The main purpose of normalization is to remove correlations with the absolute intensity of the total PAH emission and hence variations related to the PAH abundance (and distance in studies containing sources at varying distances). Here we discuss first the effects of extinction and dereddening on the observed correlations for the whole map, subsequently we present the correlations subdivided by the individual regions discussed earlier. In each case the points represent the independent pixels from the 2$\times$2 binned cubes and correlation coefficients are Pearson weighted correlation coefficients. The uncertainties were calculated from the signal to noise estimates described in Section~\ref{sec:decomp}.  

\begin{figure*}
	\begin{center}
	\includegraphics[width=16cm]{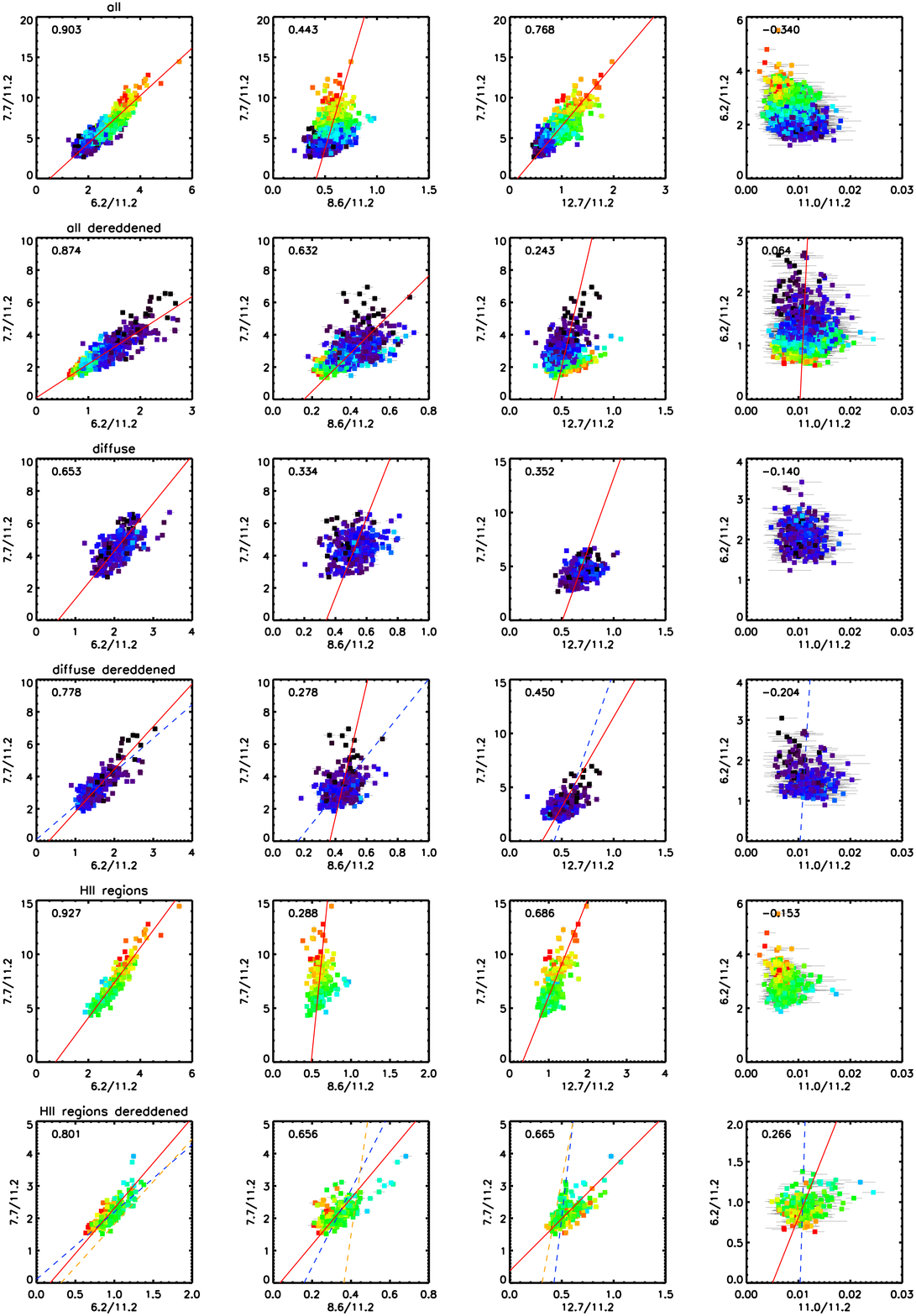}
	\end{center}
	\caption{Main correlation plots for entire W49A map. The top two rows present the correlation plots using the entire map as observed, and those derived from the dereddened map. This pattern is then repeated for the diffuse points and the points associated with the UC-\HII\ regions. Points are color coded according to their measured silicate absorption from red ($\tau_{9.8}$ $\sim$ 5) to black ($\tau_{9.8}$ $\sim$ 1). The solid red lines represent the best fits to the data and the number in the top left corner is the weighted correlation coefficient for these points. The blue dashed lines in the lower panels are the same as the dereddened correlations for the whole sample such that changes in gradient are easier to see. Additionally in the \HII\ regions sample we have included an orange dashed line representing the correlation found for the diffuse material only.}
	\label{fig2}
\end{figure*}

The main correlation plots resulting from the analysis of the whole W49A map are presented in the top two rows of Figure~\ref{fig2}, with the top row representing the observed data and the second row showing the dereddened data. 

\paragraph{7.7 vs 6.2} The $I_{6.2}$/$I_{11.2}$ vs $I_{7.7}$/$I_{11.2}$ correlation is recovered as expected, with one of the highest correlation coefficients in the entire sample at 0.903. The quality of the correlation is lowered somewhat after dereddening as shown by the correlation coefficient decreasing to 0.874. This is likely due an increase in the scatter of the points because the uncertainties associated with the extinction measurement are propagated through to the band measurements. In terms of the properties of this correlation, dereddening alters the gradient and $y$-intercept such that these parameters approach those found by previous studies (Table~\ref{table:PAHparms}). The gradient of the $I_{6.2}$/$I_{11.2}$ vs $I_{7.7}$/$I_{11.2}$ correlation falls from 2.94 in the observed data to 2.08 after dereddening. This change is accompanied by the $y$-intercept approaching zero. 

\paragraph{7.7 vs 8.6} For the $I_{7.7}$/$I_{11.2}$ vs $I_{8.6}$/$I_{11.2}$ relationship the points are much more scattered, to the point where it is hard to claim a correlation. This seems to be because the high extinction points behave very differently from the low extinction points for W49A. The $I_{8.6}$/$I_{11.2}$ ratio is less affected by extinction because both bands are afflicted by approximately the same extinction, conversely the $I_{7.7}$/$I_{11.2}$ is heavily affected by extinction because the 11.2 \micron\ band is much more heavily extinguished than the 7.7 \micron\ band. This effectively stretches the high extinction points away from the correlation that is usually recovered. Following the extinction correction the correlation coefficient is slightly increased, but still low. This is mainly because the cloud of low extinction points from the periphery of W49A, for which there is no correlation as we shall see later, dominate numerically over the extinction corrected points for the UC~\HII\ regions which do form a correlation.

\paragraph{7.7 vs 12.7} A similar problem affects the $I_{7.7}$/$I_{11.2}$ vs $I_{12.7}$/$I_{11.2}$ correlation where the extinction corrected correlation seems to split into two overlapping branches, representing the central parts of W49A near UC~\HII\ regions (light points) and the fainter regions on the periphery of W49A. In contrast to the $I_{7.7}$/$I_{11.2}$ vs $I_{8.6}$/$I_{11.2}$ ratio though, the gradient of the diffuse points for $I_{7.7}$/$I_{11.2}$ vs $I_{12.7}$/$I_{11.2}$ is similar to that measured for other objects, albeit at a low correlation coefficient. The fact that the \HII\ region points in the $I_{7.7}$/$I_{11.2}$ vs $I_{12.7}$/$I_{11.2}$ correlation plot have a different slope to the diffuse points is somewhat unexpected. We have investigated this phenomenon in terms of possible systematics associated with the decomposition method, initially focusing on possible contamination from the adjacent [Ne~{\sc ii}] emission line (see Section~\ref{sec:decomp}) and moving on to examining the local continuum for the affected spectra as these points also have a strong rising continuum in contrast to the diffuse points. Individual examination of the affected points revealed that there was no systematic offset arising. As such, the presence of a branch representing the ionized points remains a mystery -- most likely an unresolved systematic. However, such differences could fit with previous suggestions about the nature of the 12.7 \micron\ band and its relationship with edge structure (e.g. \citealt{2001A&A...370.1030H}). The different behavior of the band in the centers of UC-\HII\ regions could then reflect the enhanced PAH destruction in these regions which would potentially lead to more ragged edge structures and increased 12.7 \micron\ emission.

\paragraph{6.2 vs 11.0} Finally, $I_{6.2}$/$I_{11.2}$ vs $I_{11.0}$/$I_{11.2}$, displays a very interesting property: there appears to be no variation in $I_{11.0}$/$I_{11.2}$ to accompany the large changes in $I_{6.2}$/$I_{11.2}$. In fact, $I_{11.0}$/$I_{11.2}$ adopts a value of around 0.01 before and after dereddening with the dereddening process having little effect on this band ratio because the bands are very close in wavelength. This is contrary to expectations as laboratory studies (e.g. \citealt{1999ApJ...516L..41H}) and observations (e.g. \citealt{2011A&A...532A.128R}) have found that the 11.0 \micron\ band traces PAH ionization very well. Previous measurements of the $I_{11.0}$/$I_{11.2}$ band ratio are in the ranges of around 0.02 -- 0.04 (e.g. Peeters et al. 2013 in prep) to 0.05 -- 0.20 (e.g. \citealt{2011A&A...532A.128R, 2012A&A...537C...5R}), which exceed that seen for W49A by factors of at least 2 and 5 respectively. It is therefore puzzling that large fluctuations in the main bands are not accompanied by variations in the strength of the 11.0 \micron\ band for W49A.

\paragraph{Differences between diffuse and ionized regions} In the subsequent rows of Figure~\ref{fig2} we show the same correlation plots, except separating the points associated with the outskirts of W49A (rows 3 and 4 for observed and dereddened data respectively) and those associated with the UC-\HII\ regions in the centre. It is clear from these points that the behaviour of the two components is very different in terms of their PAH band correlations. For the diffuse regions, the dereddening process has only a small effect on the distribution of points in the correlation points. Only the 6.2 and 7.7 \micron\ bands display any sign of a correlation, with the other correlations being essentially a cloud of points around a specific point. For example, the $I_{7.7}$/$I_{11.2}$ vs $I_{8.6}$/$I_{11.2}$ correlation takes the form of a cloud of points at approximately $I_{7.7}$/$I_{11.2}$ $\simeq$ 3 and $I_{8.6}$/$I_{11.2}$ $\simeq$ 0.45. In contrast, the UC-\HII\ region points are affected by the dereddening process. As with the total sample, it is the  $I_{7.7}$/$I_{11.2}$ vs $I_{8.6}$/$I_{11.2}$ correlation which appears to be the most affected, in that a correlation appears after dereddening. However, there are systematics appearing in the distribution of the points.  The points are not evenly spread with respect to their extinction, with the highest extinction points remaining clustered away from the center of the distribution. This divergence is likely because the extinction correction method assumes that silicate absorption and PAH emission are entirely spatially separate. It is more likely that some fraction of the silicate extinction is mixed within the PAH emitting material. There is some evidence for this in that the areas of highest extinction in Figure~\ref{fig:ext} appear loosely related to the emitting regions, i.e. they obscure the centre of W49A preferentially and there are no areas of high extinction on the periphery.

\begin{figure*}
	\begin{center}
	\includegraphics[width=16cm]{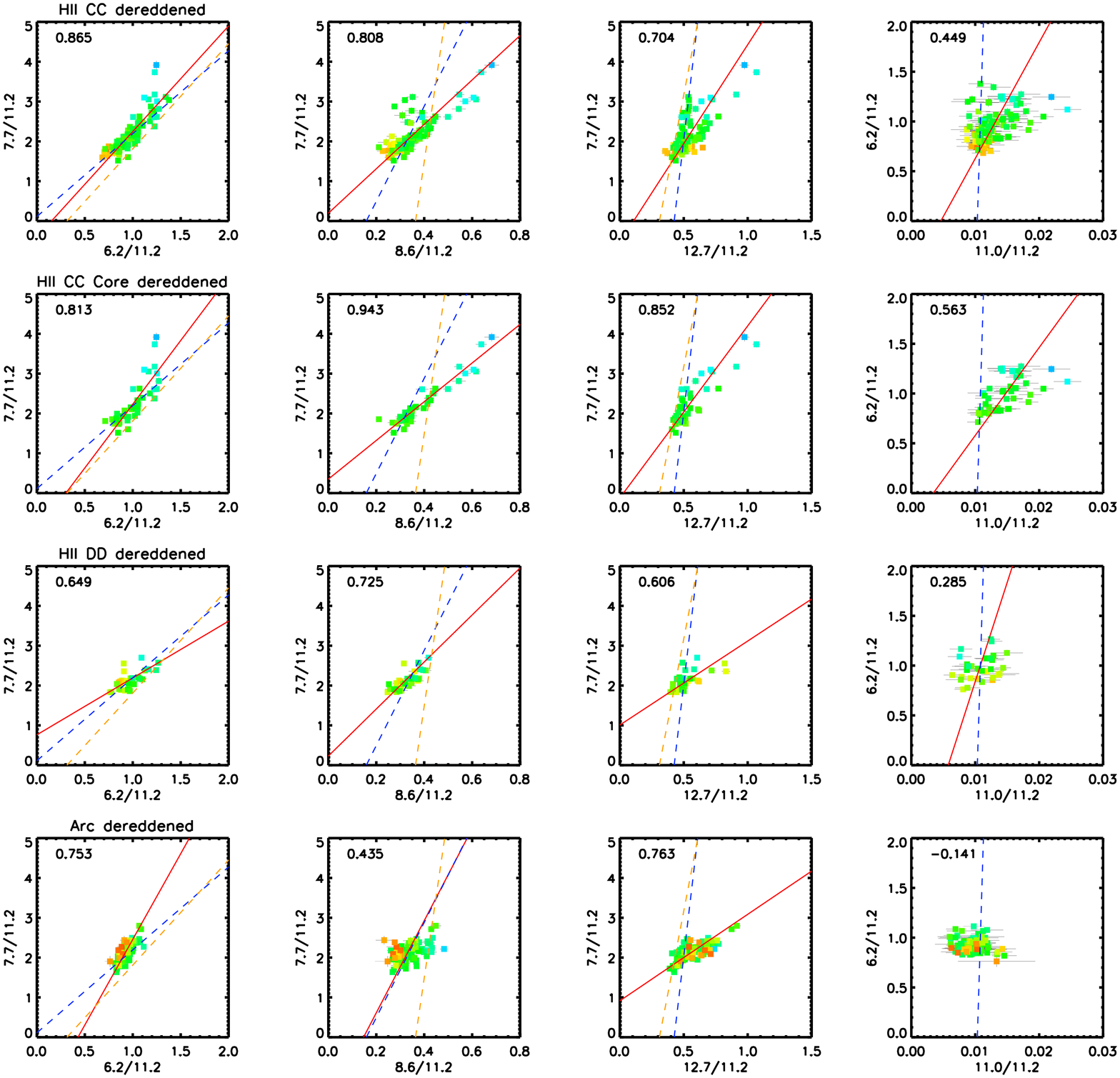}
	\end{center}
	\caption{Main correlation plots for the individual regions within W49A. Each row presents the dereddened correlation plots for specific areas of W49A. From top to bottom: CC; CC core; DD; Arc. For each combination of bands, the axis scales have been preserved to enable easy comparisons of the ranges and gradients of the correlations. As in Figure~\ref{fig2}, points are color coded according to their measured silicate absorption from red ($\tau_{9.8}$ $\sim$ 5) to black ($\tau_{9.8}$ $\sim$ 0.5). Best fit lines for each correlation are included as red solid lines, along with the weighted correlation coefficient. The blue and orange dashed lines are the same as the dereddened correlations for the whole sample (blue; Figure~\ref{fig2}, second row) and diffuse regions (orange; Figure~\ref{fig2}, fourth row) such that changes in gradient are easier to see.}
	\label{fig2a}
\end{figure*}

\paragraph{Summary} The properties of the correlations for those shown in Figure~\ref{fig2}, along with those of the individual UC-\HII\ regions shown in Figure~\ref{fig2a}, are given in Table~\ref{table:PAHparms}. We note that for UC-\HII\ region CC, there is a sub-population of points which agree with the overall correlation for the complete sample of points, particularly for the $I_{7.7}$/$I_{11.2}$ vs $I_{12.7}$/$I_{11.2}$ correlation. These points are associated with the eastern section of UC-\HII\ region CC, away from the main ionized part. In general though, these correlation parameters agree with previous works

\begin{table*}
\caption{\label{table:PAHparms}Parameters of PAH Intensity Ratio Correlations in W49A regions}
\begin{center}
\begin{tabular}{c c c p{2cm} c c p{2cm} }
 & \multicolumn{3}{c}{Observed} & \multicolumn{3}{c}{Dereddened} \\
Region & Intercept & Slope &Correlation Coefficient& Intercept & Slope & Correlation Coefficient\\
\hline
\\
\multicolumn{7}{l}{$I_{6.2}$/$I_{11.2}$ vs $I_{7.7}$/$I_{11.2}$}\\
             All &   -1.51$\pm$    0.09 &    2.94 $\pm$   0.04 &   0.903 &    0.11 $\pm$   0.03 &    2.08 $\pm$   0.03 &   0.874 \\         
         Diffuse &   -1.63$\pm$    0.33 &    2.96 $\pm$   0.17 &   0.653 &   -0.83 $\pm$   0.18 &    2.64 $\pm$   0.12 &   0.778 \\
UC~\HII\ regions &   -2.38$\pm$    0.24 &    3.25 $\pm$   0.08 &   0.927 &   -0.49 $\pm$   0.12 &    2.80 $\pm$   0.13 &   0.801 \\
             Arc &   -2.16$\pm$    0.25 &    3.23 $\pm$   0.09 &   0.967 &   -1.85 $\pm$   0.33 &    4.30 $\pm$   0.36 &   0.753 \\
              CC &   -3.01$\pm$    0.56 &    3.36 $\pm$   0.20 &   0.865 &   -0.41 $\pm$   0.15 &    2.65 $\pm$   0.16 &   0.865 \\
         CC Core &   -3.46$\pm$    1.52 &    3.51 $\pm$   0.57 &   0.647 &   -0.97 $\pm$   0.34 &    3.20 $\pm$   0.35 &   0.813 \\
              DD &   -3.53$\pm$    1.30 &    3.57 $\pm$   0.44 &   0.821 &    0.76 $\pm$   0.25 &    1.43 $\pm$   0.26 &   0.649 \\
\\
IRAS 12063-6259$^a$ & -0.60$\pm$0.11 & 3.19$\pm$0.06 & 0.961 &  0.53$\pm$0.07 & 2.27$\pm$0.06  & 0.947\\
NGC 2023$^b$ & -0.29$\pm$0.03 &  2.04$\pm$0.02  & 0.967  \\
\\

\multicolumn{7}{l}{$I_{8.6}$/$I_{11.2}$ vs $I_{7.7}$/$I_{11.2}$}\\
             All &  -17.92 $\pm$   1.35 &   43.16 $\pm$   2.46 &   0.443 &   -1.89 $\pm$   0.14 &   11.93 $\pm$   0.38 &   0.632 \\
         Diffuse &   -8.33 $\pm$   2.00 &   24.44 $\pm$   3.83 &   0.334 &  -14.97 $\pm$   3.96 &   41.15 $\pm$   8.96 &   0.278 \\
UC~\HII\ regions &  -35.21 $\pm$   8.64 &   72.01 $\pm$  14.69 &   0.288 &   -0.25 $\pm$   0.18 &    7.18 $\pm$   0.56 &   0.656 \\
             Arc &  -23.41 $\pm$   5.39 &   52.14 $\pm$   9.31 &   0.470 &   -1.70 $\pm$   0.89 &   11.56 $\pm$   2.68 &   0.435 \\
              CC & -100.62 $\pm$ 157.03 &  179.77 $\pm$ 262.52 &   0.190 &    0.20 $\pm$   0.15 &    5.58 $\pm$   0.43 &   0.808 \\
         CC Core &    1.53 $\pm$   0.53 &    7.15 $\pm$   0.84 &   0.792 &    0.35 $\pm$   0.11 &    4.86 $\pm$   0.30 &   0.943 \\
              DD &   -9.92 $\pm$   4.74 &   28.70 $\pm$   7.91 &   0.599 &    0.23 $\pm$   0.33 &    5.90 $\pm$   1.01 &   0.725 \\

\\
IRAS 12063-6259$^a$ & -1.71 $\pm$ 0.19 & 5.24 $\pm$ 0.15 & 0.954 & -12.77 $\pm$ 3.17 & 17.01 $\pm$ 3.37 & 0.636\\
NGC 2023$^b$ &   1.17$\pm$0.03 & 4.40$\pm$0.06 & 0.953 \\
\\

\multicolumn{7}{l}{$I_{12.7}$/$I_{11.2}$ vs $I_{7.7}$/$I_{11.2}$}\\
             All &   -1.12 $\pm$   0.14 &    7.62 $\pm$   0.16 &   0.768 &   -7.10 $\pm$   0.69 &   20.46 $\pm$   1.49 &   0.353 \\
         Diffuse &  -13.52 $\pm$   2.62 &   26.70 $\pm$   3.89 &   0.352 &   -4.61 $\pm$   0.74 &   16.31 $\pm$   1.54 &   0.465 \\
UC~\HII\ regions &   -2.92 $\pm$   0.56 &    9.01 $\pm$   0.52 &   0.686 &    0.14 $\pm$   0.13 &    4.16 $\pm$   0.27 &   0.640 \\
             Arc &   -1.82 $\pm$   0.62 &    7.51 $\pm$   0.56 &   0.766 &    0.86 $\pm$   0.10 &    2.58 $\pm$   0.21 &   0.764 \\
              CC &   -3.93 $\pm$   1.03 &   10.19 $\pm$   0.98 &   0.686 &   -1.21 $\pm$   0.37 &    7.16 $\pm$   0.79 &   0.678 \\
         CC Core &    0.07 $\pm$   0.74 &    6.03 $\pm$   0.75 &   0.726 &   -0.80 $\pm$   0.38 &    6.35 $\pm$   0.81 &   0.788 \\
              DD &    0.64 $\pm$   0.68 &    5.99 $\pm$   0.63 &   0.833 &    0.82 $\pm$   0.26 &    2.87 $\pm$   0.56 &   0.598 \\
\\
NGC 2023$^b$    & -1.29 $\pm$ 0.09 & 14.57 $\pm$ 0.26 & 0.904\\ 
\end{tabular}
\bigskip

$^a$: \citet{2013ApJ...771...72S}\\
$^b$: Peeters et al., 2013 in prep. NGC 2023 possesses very little extinction so we quote only the observed values.\\

\end{center}
\end{table*}

\section{Discussion}\label{sec:discuss}

\subsection{Is W49A a good prototype for extragalactic star formation?}\label{sec:sfr} 
The PAH emission from external galaxies is commonly used as a proxy for star formation (e.g. \citealt{2004ApJ...613..986P}; \citealt{2006A&A...446..877M}; \citealt{2006ApJ...653.1129B} and references therein). The PAH emission traces star formation because the copious production of UV radiation from young massive stars is the most efficient mechanism for exciting emission of the IR bands from PAH molecules. The most intense sites of star formation in the universe are the so called `starburst' galaxies in which the star formation rate may exceed many thousands of solar masses per year (while the Milky Way is thought to produce roughly one solar mass per year). The star forming regions of these galaxies cannot be spatially resolved, but are suspected to be comprised of a multitude of enormous star formation complexes like W49A, which later become super star clusters (e.g. \citealt{1995ApJ...446L...1O}; \citealt{2003ApJ...599..193F}). 

\citet{2004ApJ...613..986P} investigated the use of various diagnostic diagrams, most of which had been designed to separate active galactic nuclei (AGN) and starburst galaxies. However, they also included galactic \HII\ regions in their sample to investigate the relationship between \HII\ regions and starburst galaxies.  First we will investigate the appearance of W49A when we consider its overall appearance (the total resulting spectrum of all of the pixels in the map) to see whether it resembles a starburst or an \HII\ region. Subsequently, we will investigate the spatial scales on which the emission from the UC~\HII\ regions is diluted.

\paragraph{Integrated Properties} In order to create a comparable spectrum to extragalactic objects, every pixel of the W49A cube was coadded to create an integrated spectrum. This process included all the the emission from the non-ionized regions of W49A as clearly in an extragalactic system similar regions will be included in the beam of every observation.

Various diagnostic diagrams have been developed to help interpret such spectra in terms of the spectral differences between starburst galaxies and AGN with the goal of distinguishing between them. The most suited to the W49A observations, presented by \citet{2000A&A...359..887L}, compares the 6.2 \micron\ PAH feature strength and the 14-15 \micron\ continuum emission where both are normalized to the 5.5 \micron\ continuum. The Laurent diagnostic diagram shows the relationship between three different types of model (or template) MIR spectra: AGN, PDRs and \HII\ regions. The models used to create this diagnostic diagram were updated by \citet[][their Figure~9]{2004ApJ...613..986P}.  \citealt{2004ApJ...613..986P} found that the starburst galaxies fell midway between the \HII\ region and PDR templates. The spectrum of W49A falls at the approximate center of the cloud of starburst galaxies, regardless of whether the spectrum was dereddened or not.

A diagnostic diagram which can separate the \HII\ regions and starbursts was presented by \citet{2004ApJ...613..986P} as part of their study correlating PAH emission characteristics with the physical properties of the objects: the relationship between the number of Lyman continuum photons ($N_{Lyc}$) and the luminosity of the 6.2 \micron\ PAH emission (L$_{6.2}$). This diagram shows a clear positive linear relationship with a gap between the starbursts and the \HII\ regions. The integrated spectrum for the observed section of W49A falls on this correlation between the starbursts and \HII\ regions, albeit close to the \HII\ region regime. We estimated $N_{Lyc}$ as the sum of that found for all the \HII\ regions in the field of view by \citet{1997ApJ...482..307D}. The W49A observations only covered the northern parts of the object, and as such the total 6.2 \micron\ PAH luminosity found and the $N_{Lyc}$ figures represent only a small fraction of the true output of W49A. In fact, it has been estimated that the object contains $\sim$ 100 massive O stars giving it a total $N_{Lyc}$ figure closer to $10^{51}$ rather than the $3 \times 10^{49}$ found for the subregion studied. These numbers would firmly place W49A between the starbursts and the \HII\ regions on Figure~12 of \citet{2004ApJ...613..986P}. Furthermore the UC~\HII\ regions of W49A follow closely the population of \HII\ regions. For example, the properties of region CC (L$_{6.2}$ $\simeq$ 28 L$_\odot$; log($N_{lyc}$) = 48.9) place it well within the cloud of \HII\ region data points.

\paragraph{Spatial Dilution Effects} Given that the integrated spectrum of W49A is reminiscent of a starburst galaxy, it is of interest to see how this spectral appearance is constructed from the building blocks that make up a star forming region: UC~\HII\ regions, PDRs and molecular cloud material. Here we explore the transition between these regimes in terms of the contribution of the various components.

\begin{figure}
	\begin{center}
	\includegraphics[width=7.5cm]{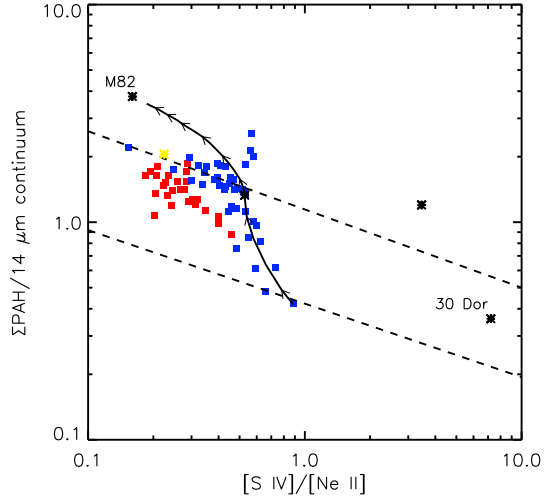}
	\end{center}
	\caption{The relationship between [S~{\sc iv}]/[Ne~{\sc ii}] and $\Sigma$PAH/VSG. Blue points represent the areas of valid data associated with UC~\HII\ region CC, red points represent the valid points from all other regions. A log scale has been adopted to allow easy comparison with Figure~14 of \citet{2006A&A...446..877M}. In addition we have included the strip which \citet{2006A&A...446..877M} found encompassed their data (dashed black lines) and the values for star forming galaxies and 30 Dor from the same study. The black arrows represent considering the integrated spectrum of larger regions around the centre of UC~\HII\ region CC and thus show the spectral dilution of the \HII\ region spectrum into the surrounding PDR-like spectra. }
	\label{figion}
\end{figure}

\citet{2006A&A...446..877M} found that there exists a relationship between an ionization proxy line ratio, [Ne~{\sc iii}]/[Ne~{\sc ii}], and the ratio of the total PAH emission ($\Sigma$PAH) to continuum emission around 14 \micron, which is usually taken to represent classical very small grains (VSG). Additional data, specific to starburst galaxies, was published by \citet{2011ApJ...728...45L}. These authors found a clear separation between the starburst galaxies and giant \HII\ regions like 30 Doradus. 

In Figure~\ref{figion}, the relationship between [S~{\sc iv}]/[Ne~{\sc ii}] and $\Sigma$PAH/VSG is shown for W49A. The non-W49A points from \citet{2006A&A...446..877M} have been converted from [Ne~{\sc iii}]/[Ne~{\sc ii}] to [S~{\sc iv}]/[Ne~{\sc ii}]  using the relationship supplied by \citet{2008MNRAS.391L.113G}. In addition, the strip indicated by \citet{2006A&A...446..877M} as the best fit to their data has also been transformed into [S~{\sc iv}]/[Ne~{\sc ii}] coordinates and encompasses the W49A points. Also included are a subset of objects for which \citet{2006A&A...446..877M} quoted [S~{\sc iv}] fluxes. Conveniently these extra points are appropriate for this study as they include the giant LMC star forming region 30 Dor, and a selection of star forming galaxies including M82. 

The W49A points are consistent with those of \citet{2006A&A...446..877M} and are particularly similar to the pattern of the spatially resolved M17 data points presented in Figure~16 of \citet{2006A&A...446..877M} in that they possess a much steeper gradient relative to the overall trend. The reason for this is clear in terms of the W49A data as the point with the lowest value of $\Sigma$PAH/VSG is associated with the center of the UC~\HII\ region CC -- the strongest 14 \micron\ continuum emission present in the W49A observations. There is a steep drop off in continuum emission with radius as distance from the UC~\HII\ regions increases. This cannot be resolved in external galaxies, confining this effect to W49A and M17 where spatial resolution is sufficient. 

To demonstrate this process, in Figure~\ref{figion} we have included a `dilution track' which has been created by taking the averaged properties of increasingly larger areas around the UC~\HII\ region CC. This track follows the path from the center pixel, which is dominated by the UC~\HII\ region (strong continuum, weaker PAH bands), to almost the whole map, which is dominated by PDR-like spectra (shallow continuum, stronger PAH bands). The track displays the steep gradient discussed earlier, with a sudden flattening of the gradient at a certain radii (R $\sim$ 8 pixels, or 0.8 pc). At that point the \HII\ region emission is diluted by the large area of PDR-like emission surrounding the \HII\ region. Interestingly, the dilution track terminates in the vicinity of the properties of M82, indicating again that the overall properties of W49A are very similar to those of M82.

In terms of the starburst / giant \HII\ region divide found by \citet{2011ApJ...728...45L}, the points for W49A straddle the boundary as might be expected from the previous discussion. Giant \HII\ regions are dominated by ionized material and as such have higher ionization and lower PAH emission. In contrast PDR dominated emission has low ionization and very high PAH emission. As shown by the dilution track, if we consider the entirety of W49A it appears as a starburst because of the copious PDR material included in the beam, while the individual UC~\HII\ regions appear as normal UC~\HII\ regions in this kind of diagnostic.

\paragraph{Outlook} It is clear that the overall properties of the W49A complex lay somewhere between the UC~\HII\ regions and starburst galaxies. Clearly the number of Lyman continuum photons for the entirety of W49A would far outstrip that corresponding the the observed area, effectively making W49A more like a starburst galaxy than we can accurately quantify here. 

Another important idea to have emerged is that the MIR appearance of W49A is only compatible with those of external starbursts if the surrounding material is included (see, for example, Figure~\ref{figion}). The difference then, between the `normal' \HII\ regions shown in Figure~12 of \citet{2004ApJ...613..986P} is the inclusion of large amounts of inter-\HII\ region material between the separate UC~\HII\ regions within W49A along with its surroundings. The divisions seen in the diagnostics can be seen as showing the divide between the observation being dominated by ionized material (e.g. local \HII\ regions) or PDR material (e.g. unresolved extragalactic star formation). In this sense, W49A represents the boundary case where the ionized regions can be resolved but are only a factor of a few greater in size than the beam size. Such arguments have relevance to the conclusions of \citet{2011ApJ...728...45L} regarding the stratification of starburst galaxies and \HII\ regions as described earlier, where the observed trends may be amplified by this simple observational effect.

\subsection{Calibrating IRAC 8 \micron\ Fluxes and PAH Emission} 

The PAH emission around 8 \micron\ is commonly used to trace star formation, its correlation with other star formation indicators has been shown by, for example, \citet{2001A&A...372..427R, 2004ApJ...613..986P, 2005ApJ...633..871C, 2005ApJ...632L..79W, 2007ApJ...666..870C}. In the era of \textit{Spitzer}, the 8 \micron\ emission is usually measured with the IRAC 8 \micron\ filter. However, the emission in the 8 \micron\ region is complex and contains multiple components -- the PAH features, the continuum and the line emission which cannot be separated in photometric observations. Here, we quantify the relative contributions in different environments and investigate the overall fraction of PAH related emission.

The average fraction of PAH, 8 \micron\ bump and 5--10 \micron\ plateau emission compared to the continuum emission in the IRAC 8 \micron\ filter can be obtained in the following manner. Firstly, convolving the IRS cube with the IRAC 8 \micron\ filter bandpass and summing the results provides a simple measurement of the flux we would expect in an IRAC 8 \micron\ observation of the same field. Subsequently, we can convolve the bandpass with the plateau fluxes and the continuum subtracted cube. This process results in measurements of the flux from continuum and PAH emission components within the IRAC filter. The ratio of the total PAH related emission to the total emission can then be easily found, along with the fractions contributed by each component. 

\begin{figure}
	\begin{center}
	\includegraphics[width=7.5cm]{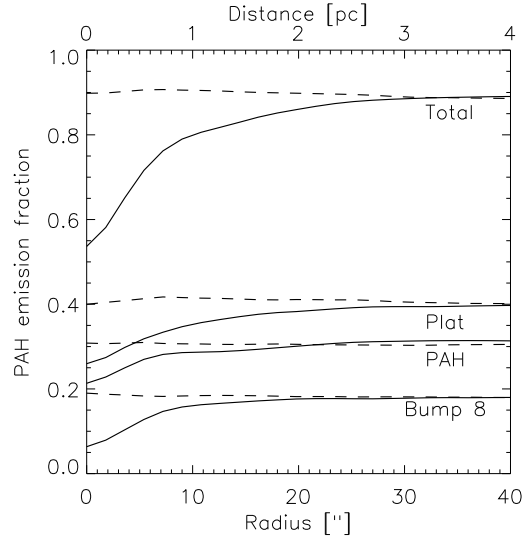}
	\end{center}
	\caption{Fraction of PAH, 5--10 \micron\ plateau and 8 \micron\ bump emission in IRAC 8 \micron\ photometry for W49A with respect to the total emission. Solid lines represent expanding regions centered upon the UC~\HII\ region CC. The low initial values are due to the strong continuum emission in the center. The dashed lines represent the results of the same process, but applied to the arc region. Each pair of lines is labeled with the corresponding emission component, including a total.}
	\label{figfrac}
\end{figure}

The total PAH related contribution is around 0.89 for the regions away from UC~\HII\ regions, and drops to around 0.55 within the UC~\HII\ regions. In Figure~\ref{figfrac} we show the effective PAH/total flux ratio for increasing radius centered on UC~\HII\ region CC and the same process centered on the arc region. Both lines in Figure~\ref{figfrac} clearly approach values of around 0.9 when approaching larger spatial regions. At lower radii though, the trends diverge, with the strong emission from the UC~\HII\ region driving the fraction of PAH emission down towards its centre. Conversely, for the arc trend the fraction of PAH emission is higher for low radii, representing the average value for pixels which are not in close proximity to the UC~\HII\ regions. Our results agree with those of \citet{2007ApJ...656..770S}, who used PAHFIT and found that the fraction of PAH emission is around 80\% in the SINGS sample of local star forming galaxies. 

This result echoes the result of Section~\ref{sec:sfr}, where it was found that spatial dilution of the UC~\HII\ spectra in the surrounding PDR emission could drastically change the spectral appearance of star formation. In this case we find that the fraction of PAH related emission is around 0.9 when more than around 2 pc from the center of UC~\HII\ region CC, but that this value drops to around 0.55 as radius decreases. 

\subsection{Novel behaviour of the 6.2 \micron\ band?}\label{sec:struct}

In earlier sections it was remarked upon that the 6.2 \micron\ band and the 7.7 \micron\ band, which are usually found to correlate extremely well, showed hints of deviation from the usual correlations. In Section~\ref{sec:pahmaps} this behavior was seen to be localized to the UC-\HII\ regions CC and DD, however it is somewhat difficult to see because of the contrast required to display the rest of the maps. Hints of this deviation also appear in Figure~\ref{fig2a} in the correlations relating to the core region of UC-\HII\ region CC. Given that the 7.7 \micron\ and 8.6 \micron\ bands correlate well in these environs, it seems that it is the 6.2 \micron\ band which is displaying anomalous behavior.

\begin{figure*}
	\begin{center}
	\includegraphics[width=16cm]{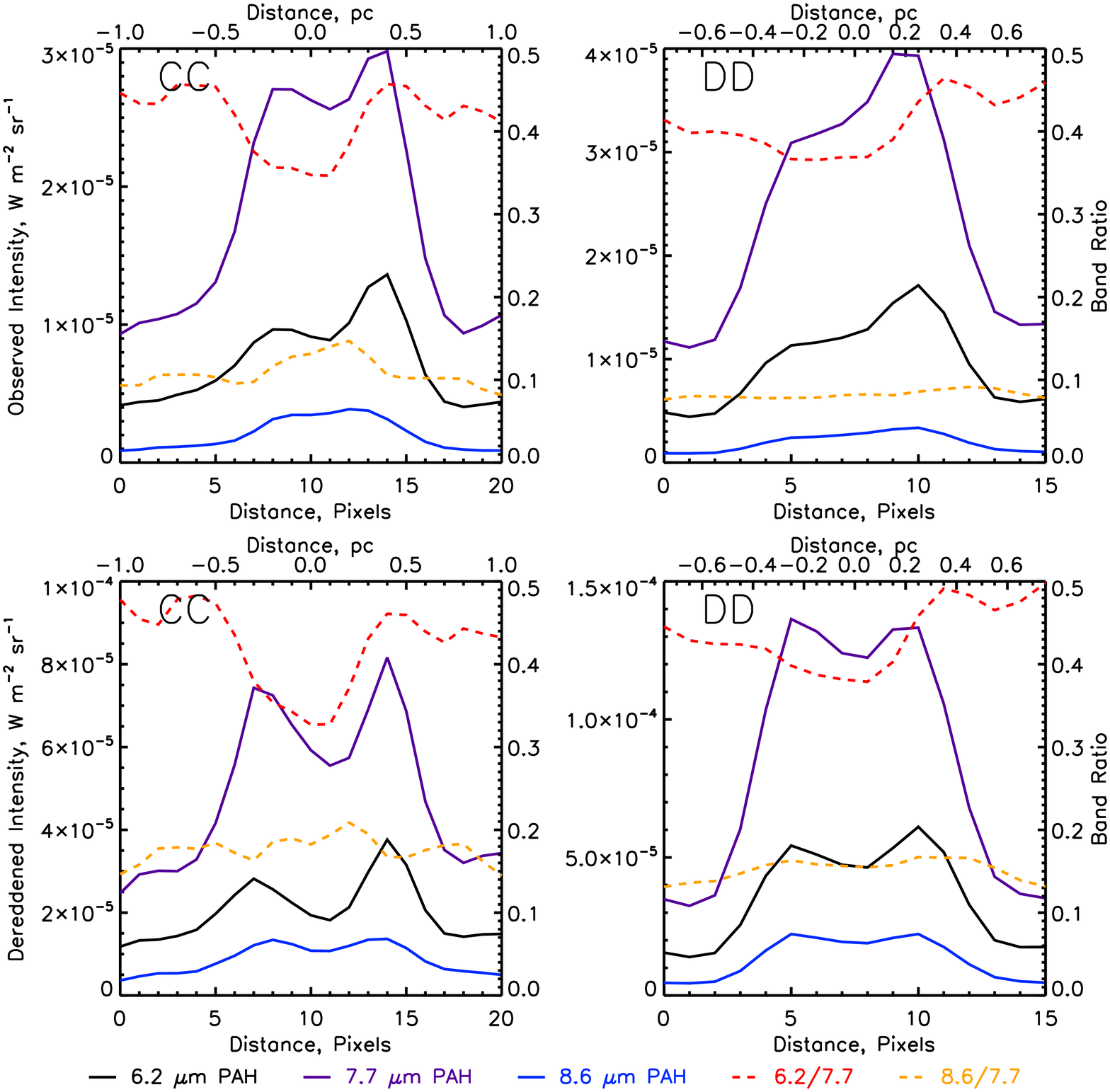}
	\end{center}
	\caption{Observed (top row) and dereddened (bottom row) intensity cuts for the 6.2, 7.7 and 8.6 \micron\ bands through the center of UC~\HII\ regions CC and DD (left and right panels respectively). Also included are the spatial profiles of the PAH band ratios for 6.2/7.7 and 8.6/7.7.  The bottom $x$ axes represent the number of pixels in the cut while the top one represent the projected distance along the cut from the center of the respective UC-\HII\ regions. }
	\label{figcutdemo}
\end{figure*}

To further investigate this effect, we first consider cuts showing the spatial profiles of the different emission components through UC-\HII\ regions CC and DD in the PAH band measurements. The resulting spatial emission profiles are shown in Figure~\ref{figcutdemo} for both the data before and after dereddening (top and bottom rows). The cut through CC is taken approximately north-south and coincident with the pixel grid of the maps, while for DD the most symmetric cuts are found for a PA:45$^\circ$. 

The general conclusions from merely observing the spatial profiles of the emission components and ratios are that for UC-\HII\ region CC: a) in the UC~\HII\ regions the 6.2 and 7.7 \micron\ PAH emission bands have similar spatial emission profiles, while the 8.6 is more symmetric; and b) that the 6.2/7.7 ratio has a strong central minimum. Conversely for UC-\HII\ region DD, the 7.7 and 8.6 bands appear to have similar spatial emission profiles while the 6.2 band differs, and there is also a less pronounced central minimum in the 6.2/7.7 ratio. For both regions the 8.6/7.7 band ratio remains approximately constant (albeit with a small bump in the centre for the CC observations before dereddening). The variations in the 6.2/7.7 band ratio exceed those seen in other objects. For example, in NGC 2023 (see Peeters et al 2014, in prep), the 6.2/7.7 band ratio assumes values between $\sim$ 0.45 -- 0.6 throughout, while the minimum for W49A/CC is $\sim$ 0.35. 

The decoupling of the 6.2 \micron\ band from those at 7.7 and 8.6 \micron\ is troubling as usually these bands are found to vary together (e.g. \citealt{2002A&A...382.1042V,2008ApJ...679..310G}). A similar effect has only been observed in one other object, N66 the giant \HII\ region in the small Magellanic cloud \citep{2013ApJ...771...16W}. In fact, the decoupling effect is visible in the dereddened correlation between 6.2/11.2 and 7.7/11.2 (Figure~\ref{fig2a}). For the CC core region the variations in the 6.2 \micron\ band cease at around I$_{6.2}$/I$_{11.2}$ = 1.3 while the I$_{7.7}$/I$_{11.2}$ ratio is still increasing. In the correlation plots this effect is only visible after dereddening because it is obscured by the high number of moderate extinction points. However, the observed and dereddened spatial profiles are very similar because the extinction at 6.2 \micron\ is very similar to that at 7.7 \micron\ in the \citet{2006ApJ...637..774C} extinction law. 

The 6.2, 7.7 and 8.6 \micron\ bands all have different vibrational assignments in the PAH model. The 6.2 \micron\ band is thought to be generated by aromatic C-C stretches, the 8.6 \micron\ band arises from C-H in plane bending modes, and the 7.7 \micron\ band is thought to originate in a combination of these vibrations. From that perspective a breakdown in the 6.2 / 7.7 relationship could be due to the C-H component of the 7.7 \micron\ band becoming dominant over the C-C component. In that case, one would expect that the $I_{7.7}$/$I_{11.2}$ vs $I_{8.6}$/$I_{11.2}$ correlation would be strengthened as the dominant vibrational mode for 7.7 would then be the same as for 8.6. This is indeed what is seen in both the spatial cuts (Figure~\ref{figcutdemo}) and the correlation plots, where a better correlation is found for $I_{7.7}$/$I_{11.2}$ vs $I_{8.6}$/$I_{11.2}$ than $I_{7.7}$/$I_{11.2}$ vs $I_{6.2}$/$I_{11.2}$ for CC core and DD.

\subsection{The ``Diffuse" Emission}\label{sec:diffem}
It is obvious that the regions of the W49A map away from the center of W49A behave differently than those associated with the ionized regions towards the center. Here, we compare the emission surrounding W49A with previous studies of diffuse PAH / mid-IR emission with the aim of determining whether the PAH emission seen surrounding W49A is associated with W49A, or rather emission from the copious material along the line of sight.

\begin{figure}
	\begin{center}
	\includegraphics[width=7.5cm]{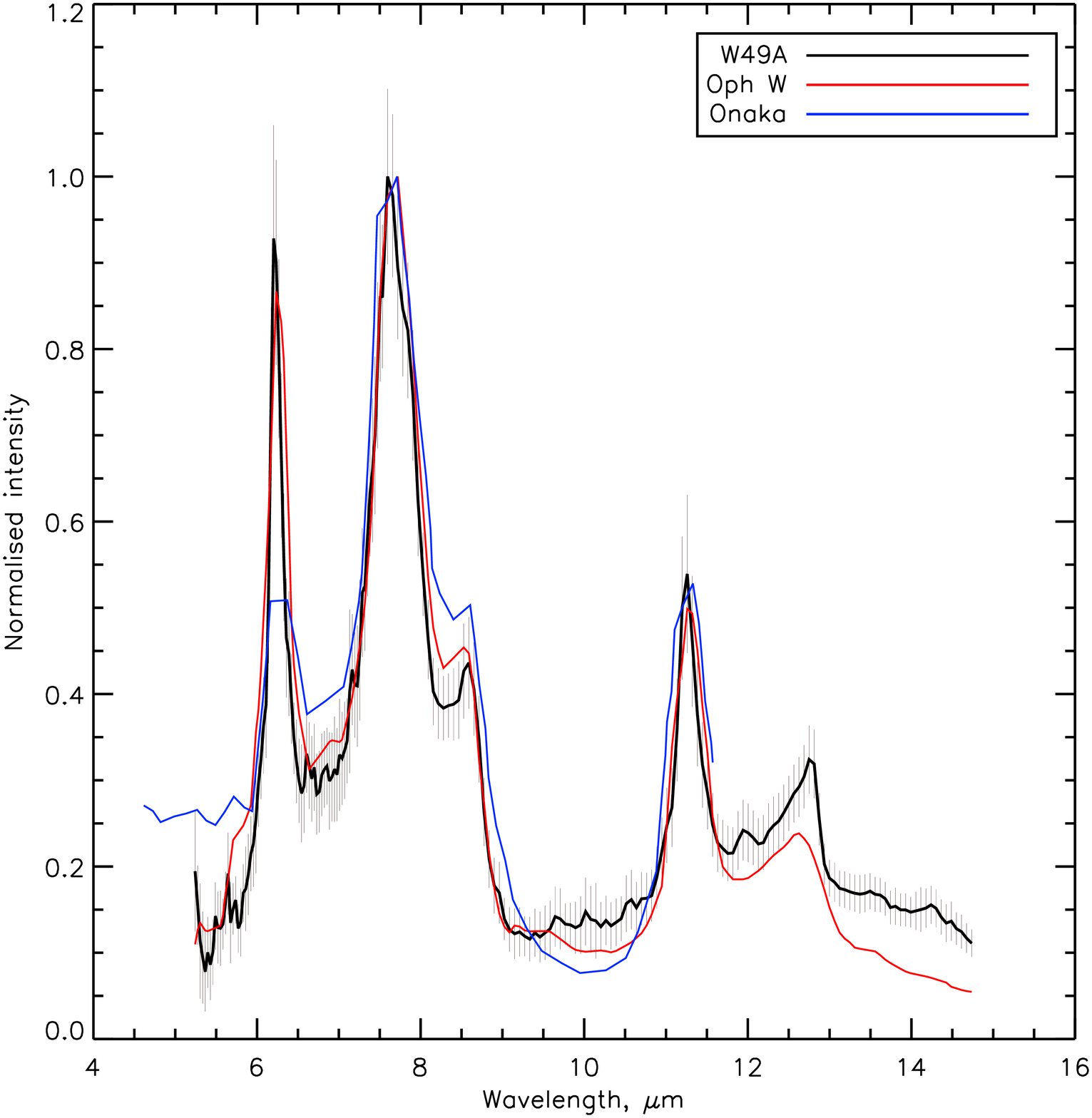}
	\end{center}
	\caption{Composite dereddened spectrum of the W49A diffuse emission (black) compared to other published spectra of diffuse emission resampled to the same spectral resolution. The spectrum of the western region of the Ophiuchus molecular cloud is given by \citet{boul_ism} and shown in red. The second region of diffuse galactic emission studied by \citet{1996PASJ...48L..59O} is shown in blue. There are main difference is that the emission at wavelengths greater than $\sim$ 12 \micron\ is much lower for the non-W49A spectra.}
	\label{figdiff}
\end{figure}

In Figure~\ref{figdiff}, an example of the `diffuse' emission around the outskirts of W49A is shown normalized to the peak intensity. This spectrum was obtained by averaging the dereddened data points for a ten pixel square region of the W49A cube (using pixels in the $x$ range 1--11 and $y$ range 15--25; shown in Figure~\ref{fig1}). We have chosen to show the dereddened version as the comparison diffuse spectra are for very low extinction sightlines. The error bars shown in Figure~\ref{figdiff} are generated by finding the statistical variance of all of the flux values for a given wavelength, rather than by combining their quoted uncertainties. This gives a better measure of the uncertainty of the points in the periphery of W49; certainly, the plotted uncertainties are larger than the rms uncertainties. The resulting uncertainties are therefore reasonably large so as to better reflect the variability of the large area coadded, and amount to roughly 15\% at each wavelength point.

Previous studies, for example \citet{1996PASJ...48L..59O, boul_ism,2003A&A...405..999K}, have focused exclusively on the properties of the diffuse emission. Comparison with the W49A `diffuse' spectrum shows that the W49A spectrum matches that observed for the Ophiuchus molecular cloud with ISOCAM reasonably well, especially in the 5--12 \micron\ regime. The agreement is weaker at longer wavelengths where the continuum is weaker relative to the W49A spectrum. The shape of the 12.7 \micron\ feature is also slightly different between the W49A spectrum and the Ophiuchus spectrum. In particular, the W49A spectrum is more sharply peaked at around 12.7 \micron\ compared to the Oph. spectra which is more rounded, however this could be due to the reduced resolution of the ISOCAM instrument used to observe the Ophiuchus spectrum. The 12.7 \micron\ feature profile found in W49A is more similar to the average PAH spectrum given by \citet{2001A&A...370.1030H}. It should be noted that there appears to be very little contamination from the 12.8 \micron\ [Ne~{\sc ii}] line in the W49A `diffuse' spectrum. The \citet{1996PASJ...48L..59O} spectrum of diffuse galactic emission obtained using the MIRS instrument \citep{1994ApJ...428..370R} on board the IRTS \citep{1994ApJ...428..354M} also differs in its 6.2 \micron\ emission as it has less than a quarter of the strength.

The discrepancy with the shape of the 12.7 \micron\ feature and the higher continuum at $\lambda > 12$ \micron\ suggests the possibility that the emission from the outskirts of W49A is not truly diffuse emission, but rather the emission associated with the molecular cloud material remaining around the edges of W49A which is being slightly illuminated by the escaping UV photons from the inner regions. A rough estimate of the emission from the vicinity of W49A can be gained by calculating the expected emission along the line of sight and subtracting it from that observed. We use the diffuse spectrum given by \citet[][Figure~5.13]{2005pcim.book.....T} as a reasonable proxy for the line of sight emission. This spectrum was scaled to the column density along the line of sight to W49A, which was measured by \citet{2012A&A...540A..87G} and found to be $N_H$ = $460 \times 10^{20}$ cm$^{-2}$ to within a factor of around two. Comparing the scaled diffuse spectrum to the W49A spectrum, it is clear that there is a significant contribution from line of sight emission. For example, the peak strength of the 6.2 \micron\ feature in the W49A `diffuse' spectrum is $\sim$ 8 $\times$ 10$^{-6}$ W m$^{-2}$ sr$^{-1}$, while the scaled diffuse spectrum yields $\sim$ 9 $\times$ 10$^{-6}$ W m$^{-2}$ sr$^{-1}$. This then suggests that the majority of the diffuse emission is emitted along the line of sight, rather than near W49A. In summary, although there is a large uncertainty of around 50\% associated with the measurement of the column density which was used to calculate the expected line of sight PAH emission, it seems very likely that the emission in the outskirts of W49A is dominated by line of sight emission.

\section{Summary and Conclusions}\label{sec:conc}
In this paper, we have created and analyzed a large \textit{Spitzer}/IRS-SL spectral cube containing observations of the galactic star forming region W49A. Using the spline method, the MIR emission from each pixel of the W49A cube was decomposed into its basic components (continuum, emission lines, PAH features). This data was then investigated from a number of perspectives. Initially, maps and correlation plots were created, suggesting a disruption of the usual PAH band correlations. Subsequent investigations then probed these behaviors in terms of the structure of the UC~\HII\ regions contained within W49A and found that the disruptions are highly localized to the shells of two such regions. The nature of the emission surrounding W49A was also investigated, along with the use of W49A as a proxy for starburst galaxies. The main conclusions drawn from these analyses are as follows:

\begin{enumerate}
\item The MIR spectra of W49A are obscured by a large amount of extinguishing material, both internal and along the line of sight. Our extinction measurements are consistent with the line of sight extinction being around 2 magnitudes in the K band, while internal extinction can account for as much as 3 additional magnitudes of K band extinction.

\item In general, the PAH correlations recovered for W49A match those found for other objects, however some UC~\HII\ regions within W49A possess much higher quality correlations than others, perhaps due to uncertainties in the extinction corrections.

\item The 6.2 \micron\ PAH band appears to be disconnected from the 7.7 and 8.6 \micron\ PAH bands in the two most prominent UC~\HII\ regions in W49A -- a phenomena which has previously only been detected in an SMC giant star forming region. This phenomenon is visible both in the maps of PAH emission and the correlation plots.

\item The MIR spectra of W49A appears reminiscent of a starburst galaxy only if the entirety of W49A is considered. Including a larger fraction of PDR / diffuse material in the beam changes the average spectrum from that of an \HII\ region to that of a starburst. The integrated properties of the field of view studied suggest that W49A exists in a regime between normal \HII\ regions and starburst galaxies.

\item The regions of `diffuse' material surrounding W49A are likely to be line of sight emission from large scale galactic structure and not associated with W49A. The spectra of this material closely matches that of regions on the periphery of other molecular clouds / star forming regions.

\item The fraction of aromatic emission in the IRAC 8 \micron\ filter for star forming regions depends on whether the source is resolved. In the total W49A spectrum (coadding every pixel), the average fraction of PAH emission from bands and plateaus in the IRAC 8 \micron\ filter is around $\simeq$~90\% -- similar to prior findings. However for smaller apertures different values can be obtained, for example a small aperture centered on the UC~\HII\ region CC provides a much lower fraction of around 55\% because the beam is dominated by the ionized region.  \\
\end{enumerate}

One of the main results of this study, that the 6.2 \micron\ PAH band decouples from those at 7.7 and 8.6 \micron\ in some circumstances, has only been observed in one object prior to this work. It is important to note that for the W49A map, this effect is only obvious when very small regions are investigated in detail -- inspection of the correlation plots for the whole map does not lead to this result. The small numbers of detected sources showing this decoupling then likely reflect the very small numbers of sources that have been investigated with this spatial detail. Of existing and past facilities, only the \textit{Spitzer}/IRS spectral mapping mode is likely to be able to discern this kind of effect and the difficulties of reduction and interpretation of a spectral cube with thousands of independent pixels have hampered such studies. Future analysis of the archival \textit{Spitzer}/IRS spectral cubes and forthcoming \textit{James Webb Space Telescope} instruments NIRSpec and MIRI will likely yield detections of similar details that will hopefully spur refinements to the PAH model.

\acknowledgments
DJS thanks the referee for helpful comments regarding the structure and discussion of this paper which have greatly improved it. DJS and EP acknowledge support from an NSERC Discovery Grant and an NSERC Discovery Accelerator Grant. WDYC acknowledges support from an NSERC Undergraduate Student Research Award. DJS and EP thank Maryvonne Gerin for discussions regarding the sub millimeter observations of W49A. This work is based on observations made with the \textit{Spitzer Space Telescope}, which is operated by the Jet Propulsion Laboratory, California Institute of Technology under a contract with NASA. This research has made use of NASA's Astrophysics Data System.

\bibliographystyle{apj}

\end{document}